\documentclass[fleqn,usenatbib]{mnras}
\usepackage{newtxtext,newtxmath}
\usepackage[T1]{fontenc}
\DeclareRobustCommand{\VAN}[3]{#2}
\let\VANthebibliography\thebibliography
\def\thebibliography{\DeclareRobustCommand{\VAN}[3]{##3}\VANthebibliography}
\usepackage{graphicx}	
\usepackage{amsmath}	
\usepackage{soul}
\usepackage{tabularx,ragged2e}
\usepackage{booktabs,array,times}
\usepackage[export]{adjustbox}
\usepackage{lscape}
\usepackage{rotating}
\usepackage{wrapfig}
\usepackage{lscape}
\usepackage{epstopdf}
\usepackage{float}
\usepackage[dvipsnames]{xcolor}

\def\CIVdblt{{\rm C~}\kern 0.1em{\sc iv}~$\lambda\lambda 1548, 1550$}
\def\NVdblt{{\rm N~}\kern 0.1em{\sc v}~$\lambda\lambda 1238, 1242$}
\def\OVIdblt{{\rm O~}\kern 0.1em{\sc vi}~$ 1031, 1037$}
\def\SVIdblt{{\rm S~}\kern 0.1em{\sc vi}~$ 933, 944$}
\def\SiIVdblt{{\rm Si~}\kern 0.1em{\sc iv}~$\lambda\lambda1394, 1403$}
\def\MgIIdblt{{\rm Mg~}\kern 0.1em{\sc ii}~$\lambda\lambda2796, 2803$}
\def\NeV{\hbox{{\rm Ne~}\kern 0.1em{\sc v}}}
\def\NeVI{\hbox{{\rm Ne~}\kern 0.1em{\sc vi}}}
\def\NeVIII{\hbox{{\rm Ne~}\kern 0.1em{\sc viii}}}
\def\OI{\hbox{{\rm O~}\kern 0.1em{\sc i}}}
\def\OII{\hbox{{\rm O~}\kern 0.1em{\sc ii}}}
\def\OIII{\hbox{{\rm O~}\kern 0.1em{\sc iii}}}
\def\OIV{\hbox{{\rm O~}\kern 0.1em{\sc iv}}}
\def\OV{\hbox{{\rm O~}\kern 0.1em{\sc v}}}
\def\OVI{\hbox{{\rm O~}\kern 0.1em{\sc vi}}}
\def\OVII{\hbox{{\rm O~}\kern 0.1em{\sc vii}}}
\def\OVIII{\hbox{{\rm O~}\kern 0.1em{\sc viii}}}
\def\NI{\hbox{{\rm N~}\kern 0.1em{\sc i}}}
\def\NII{\hbox{{\rm N~}\kern 0.1em{\sc ii}}}
\def\NIII{\hbox{{\rm N~}\kern 0.1em{\sc iii}}}
\def\NIV{\hbox{{\rm N~}\kern 0.1em{\sc iv}}}
\def\NV{\hbox{{\rm N~}\kern 0.1em{\sc v}}}
\def\NVII{\hbox{{\rm N~}\kern 0.1em{\sc vii}}}
\def\CII{\hbox{{\rm C~}\kern 0.1em{\sc ii}}}
\def\CIII{\hbox{{\rm C~}\kern 0.1em{\sc iii}}}
\def\SiII{\hbox{{\rm Si~}\kern 0.1em{\sc ii}}}
\def\SiIII{\hbox{{\rm Si~}\kern 0.1em{\sc iii}}}
\def\SII{\hbox{{\rm S~}\kern 0.1em{\sc ii}}}
\def\SIII{\hbox{{\rm S~}\kern 0.1em{\sc iii}}}
\def\SIV{\hbox{{\rm S~}\kern 0.1em{\sc iv}}}
\def\SV{\hbox{{\rm S~}\kern 0.1em{\sc v}}}
\def\SVI{\hbox{{\rm S~}\kern 0.1em{\sc vi}}}
\def\SiI{\hbox{{\rm Si~}\kern 0.1em{\sc i}}}
\def\PII{\hbox{{\rm P~}\kern 0.1em{\sc ii}}}
\def\AlII{\hbox{{\rm Al~}\kern 0.1em{\sc ii}}}
\def\AlIII{\hbox{{\rm Al~}\kern 0.1em{\sc iii}}}
\def\CaI{\hbox{{\rm Ca~}\kern 0.1em{\sc i}}}
\def\CaII{\hbox{{\rm Ca~}\kern 0.1em{\sc ii}}}
\def\CrII{\hbox{{\rm Cr~}\kern 0.1em{\sc ii}}}
\def\CII{\hbox{{\rm C~}\kern 0.1em{\sc ii}}}
\def\CIII{\hbox{{\rm C~}\kern 0.1em{\sc iii}}}
\def\CIV{\hbox{{\rm C~}\kern 0.1em{\sc iv}}}
\def\CV{\hbox{{\rm C}\kern 0.1em{\sc v}}}
\def\MgX{\hbox{{\rm Mg}\kern 0.1em{\sc x}}}
\def\MgI{\hbox{{\rm Mg}\kern 0.1em{\sc i}}}
\def\MgII{\hbox{{\rm Mg}\kern 0.1em{\sc ii}}}
\def\FeII{\hbox{{\rm Fe~}\kern 0.1em{\sc ii}}}
\def\FeIII{\hbox{{\rm Fe~}\kern 0.1em{\sc iii}}}
\def\H{\hbox{{\rm H~}}}
\def\HI{\hbox{{\rm H~}\kern 0.1em{\sc i}}}
\def\HeI{\hbox{{\rm He~}\kern 0.1em{\sc i}}}
\def\HII{\hbox{{\rm H~}\kern 0.1em{\sc ii}}}
\def\Lya{\hbox{{\rm Ly}\kern 0.1em$\alpha$}}
\def\Lyb{\hbox{{\rm Ly}\kern 0.1em$\beta$}}
\def\Lyg{\hbox{{\rm Ly}\kern 0.1em$\gamma$}}
\def\Lyth{\hbox{{\rm Ly}\kern 0.1em$\theta$}}
\def\Lyfive{\hbox{{\rm Ly}\kern 0.1em$5$}}
\def\Lysix{\hbox{{\rm Ly}\kern 0.1em$6$}}
\def\Lyseven{\hbox{{\rm Ly}\kern 0.1em$7$}}
\def\Lyeight{\hbox{{\rm Ly}\kern 0.1em$8$}}
\def\Lynine{\hbox{{\rm Ly}\kern 0.1em$9$}}
\def\Lyten{\hbox{{\rm Ly}\kern 0.1em$10$}}
\def\MnII{\hbox{{\rm Mn~}\kern 0.1em{\sc ii}}}
\def\kms{\hbox{km~s$^{-1}$}}
\def\cmsq{\hbox{cm$^{-2}$}}
\def\cc{\hbox{cm$^{-3}$}}

\newcommand{\angstrom}{\mbox{\normalfont\AA}}

\title[pLLS at $z \approx 0.8$]{A partial Lyman limit system tracing intragroup gas at $z \approx 0.8$ towards HE~$1003+0149$}
\author[Narayanan et al.]{Anand Narayanan$^{1}$\thanks{E-mail: anand@iist.ac.in}, Sameer$^{2}$, Sowgat Muzahid$^{3,4}$, Sean D. Johnson$^{5}$, Purvi Udhwani$^{1}$, \newauthor Jane C. Charlton$^{2}$, Valentin Mauerhofer$^{6,7}$, Joop Schaye$^8$, Mathin Yadav$^{1}$,\\
$^{1}$Department of Earth and Space Sciences, Indian Institute of Space Science \& Technology, Thiruvananthapuram 695547, Kerala, INDIA\\
$^{2}$Department of Astronomy \& Astrophysics, The Pennsylvania State University, 525 Davey Laboratory
University Park, PA, 16802, USA\\
$^{3}$Inter-University Centre for Astronomy \& Astrophysics, Post Bag 4, Ganeshkhind, Savitribai Phule Pune University Campus, Pune 411 007, India \\
$^{4}$Leibniz-Institut fuÜr Astrophysik Potsdam (AIP), An der Sternwarte 16, D-14482 Potsdam, Germany\\
$^{5}$Department of Astronomy, 1085 S. University, 323 West Hall, Ann Arbor, MI 48109-1107, USA\\ 
$^{6}$Observatoire de Gen\'eve, Universit\'e de Gen\'eve, Chemin Pegasi 51, 1290 Versoix, Switzerland \\
$^7$ Univ Lyon, Univ Lyon1, ENS de Lyon, CNRS, Centre de Recherche Astrophysique de Lyon UMR5574, 69230 Saint-Genis-Laval, France \\
$^{8}$ Leiden Observatory, Leiden University, P.O. Box 9513, 2300 RA Leiden, the Netherlands.
}
\begin{document}
\label{firstpage}
\pagerange{\pageref{firstpage}--\pageref{lastpage}}
\maketitle

\begin{abstract}

    We present analysis of the galaxy environment and physical properties of a partial Lyman limit system at $z = 0.83718$ with {\HI} and metal line components closely separated in redshift space ($|\Delta v| \approx 400$~{\kms}) towards the background quasar HE~$1003+0149$. The $HST$/COS far-ultraviolet spectrum provides coverage of lines of oxygen ions from {\OI} to {\OV}. Comparison of observed spectral lines with synthetic profiles generated from Bayesian ionization modeling reveals the presence of two distinct gas phases in the absorbing medium. The low-ionization phase of the absorber has sub-solar metallicities ($\sim 1/10$~solar) with indications of [C/O] $< 0$ in each of three components. The {\OIV} and {\OV} trace a more diffuse higher-ionization medium with predicted {\HI} column densities that are $\approx 2$~dex lower. The quasar field observed with $VLT$/MUSE reveals three dwarf galaxies with stellar masses of $M^* \sim 10^{8} - 10^{9}$~M$_\odot$, and with star formation rates of $\approx 0.5 - 1$~M$\odot$~yr$^{-1}$, at projected separations of $\rho/R_{\mathrm{vir}} \approx 1.8 - 3.0$ from the absorber. Over a wider field with projected proper separation of $\leq 5$~Mpc and radial velocity offset of $|\Delta v| \leq 1000$~{\kms} from the absorber, 21 more galaxies are identified in the $VLT$/VIMOS and Magellan deep galaxy redshift surveys, with 8 of them within $1$~Mpc and $500$~{\kms}, consistent with the line of sight penetrating a group of galaxies. The absorber presumably traces multiple phases of cool ($T \sim 10^4$~K) photoionized intragroup medium. The inferred [C/O] $< 0$ hint at preferential enrichment from core-collapse supernovae, with such gas displaced from one or more of the nearby galaxies, and confined to the group medium. 
\end{abstract}




\section{Introduction}

The gaseous halos of galaxies and the regions where they connect with the filaments of the intergalactic medium hold substantial baryonic mass in the form of metals and {\HI}, comparable to the gaseous disks of galaxies. These regions bear the imprint of large-scale gas flows in and out of galaxies which in turn regulates galaxy evolution. Among the different classes of quasar absorbers, the partial Lyman limit and Lyman limit systems (pLLS and LLS) trace halo gas bound to galaxies and in the extended unbound medium referred to as the circumgalactic medium  \citep[CGM;][]{Shull_2014}. These absorption systems, are defined to be partially or fully optically thick at the Lyman limit, which corresponds to neutral gas column densities of $16 \leq \log~[N(\HI)/\cmsq] \leq 17.2$, and $\log~[N(\HI)/\cmsq] \geq 17.2$ respectively. Recently, several authors have also highlighted the importance of pLLS and LLS as a means to discern inflows and outflows from galaxies. \citet{Lehner_2013} found a bimodality in the metallicity distribution for a sample of low redshift ($z \leq 1$) absorbers having $16.2 \leq \log [N(\HI)/\cmsq] \leq 18.5$ with distinct peaks at $\log (Z/Z_\odot)= -0.3$, and $-1.6$, and a dip at $\log (Z/Z_\odot) \approx -1$. The metal enriched portion of this sample is seen as tracing gas outflows from galaxies while the low metallicity portion is interpreted as accretion of cool gas from the IGM onto galaxies \citep{Wotta_2016}. A higher fraction of this metal-poor gas ($\log (Z/Z_\odot) < -1.4$) was found to be associated with pLLS ($16.2 \leq \log [N(\HI)/\cmsq] < 17.2$) than with LLS \citep{wotta2018metallicity}. Simulations also show cold accretion streams of low-metallicity photoionized gas contributing to the population of LLS \citep{fumagalli2011absorption,vancoldaccretion,vandevoortschaye2012}, although the bimodality in metallicity distribution, or a trend between metallicity and radial inflow - outflow kinematics is not always reproduced \citep{Hafen2017,rahmatioppenheimer2018}. Nonetheless, as \citet{2016ApJ...833..283L} point out, from an observational stand point, the high column densities ensure that even trace amounts of metals will be detected in these high {\HI} systems, which make pLLS better indicators of the lower limits on metallicities of gas outside and around galaxies compared to the clouds of the {\Lya} forest. 

The identification of pLLS and LLS with the gaseous halos of galaxies is based on the well established anti-correlation between {\HI} line strength and the impact parameter to the closest galaxy \citep[e.g.,][]{rudie2012gaseous,Rakic2012,Krogager2017,2000MNRAS.319..517L,2010MNRAS.407.1581S, steidel2010structure,2012JCAP...07..028F}, a trend which is also reproduced by simulations \citep{rahmati2014predictions}. The overall trends emerging from galaxy - absorber surveys position pLLS and LLS preferentially within $\rho \sim 100$~kpc of projected separation from the nearest detected galaxy \citep{Chen_1998, Thom_2012, Werk_2014, Johnson_2015}. More targeted studies of individual systems seen in absorption against background quasars have also reinforced their association with large-scale gas flows into and outward from galaxies \citep{Ribaudo_2013,2011Sci...334..952T}. Based on a survey of {\HI} absorption around a large sample of high-$z$ star-forming galaxies, \citet{rudie2012gaseous} conclude that many LLS at $z \approx 2 - 3$ have an origin within $300$~kpc of $M_* \sim 10^{10}$~M$_\odot$ galaxies, including fairly massive Lyman break galaxies. On the other hand, the simulations of \citet{rahmati2014predictions} show that dwarf galaxies below the detection limit  ($M* \lesssim 10^8$~M$_\odot$) are potentially the counterparts of strong {\HI} systems. Through sightlines targeted at nearby low-mass ($\log~M/M_* \sim 7.5 - 8$) dwarf galaxies, \citet{Zheng_2019, Zheng_2020} detect strong {\HI} and metal line absorption arising from their CGM. These two views are complementary since low mass galaxies tend to be heavily clustered around high-mass systems. In the ongoing CUBS survey \citet{2020MNRAS.497..498C} have identified LLS with galaxies that encompass a wide range, from star-forming to quiescent massive galaxies, dwarfs, and galaxy groups. 

It is well known that environment plays a pivotal role in shaping the distribution of gas around galaxies. Galaxy dense fields such as groups and clusters can significantly alter the covering fractions of gas in the CGM/IGM. For example, strong {\MgII} absorbers, which are statistically consistent with being drawn from the same population as pLLS, LLS and DLAs, are $> 3\sigma$ overabundant at cluster redshifts compared to the surroundings of field galaxies \citep{2008ApJ...679.1144L}. Compared to field galaxies, the covering fraction and the strength of the {\MgII} absorption are found to be higher in regions where multiple galaxies are present, hinting at environment enhancing the gas cross-section around galaxies \citep{Dutta2020}. \citet{2006MNRAS.372..959A} had reported a large clustering of {\Lya} lines in the local universe ($z \sim 0.06$) possibly connected with a large-scale intergalactic filament of gas within a group of galaxies. Complex and diverse processes that frequently unfold in over-dense fields such as star-formation and AGN-driven outflows, galaxy mergers, tidal and ram-pressure stripping of disk gas, and the accretion of IGM gas, can turn the intergalactic space within galaxy groups and clusters into gas rich environments with large {\HI} cross-section. Using quasar absorption line observations targeted at three galaxy clusters, \citet{2017ApJ...846L...8M} detected such gas with $\log~[N(\HI)/\cmsq] > 16.5$ producing partial Lyman limit absorption in the cluster outskirts ($\rho > 1.5r_{500}$). Ionization models showed these to be relatively metal-rich systems with $\log (Z/Z_\odot) \approx [-1.0, 0]$ \citep{pradeep_detection_2019}. Similarly, \citet{manuwal2019c} detected {\HI} - {\CIV} absorption tracing substantial columns of photoionized gas in the outskirts of the Virgo cluster, which was explained as interstellar gas displaced by outflows or tidal forces. Lower {\HI} column densities, in the range of  $\log~[N(\HI)/\cmsq] \approx 13.0 - 15.0$, have also been detected in cluster outskirts \citep{Yoon_2012, yoon2017lyalpha}. All these observations have opened up a new window to study the cold ($T \sim 10^4$~K) and dense ($n_{\H} \sim 10^{-3}$~{\cc}) phases of intergalactic gas within dense galaxy fields, supplementing the X-ray observations that exclusively target the hot ($T > 10^6$~K) component of the intragroup/intracluster plasma. 

In this work we present the analysis of a double-component partial Lyman limit system at $z = 0.83718$, and a $\log [N(\HI)/\cmsq] \approx 16.1$ cloud offset by $400$~{\kms} identified in the $HST$/COS spectrum of the quasar HE~$1003+0149$. The spectrum offers simultaneous information on five successive ionization stages of oxygen from {\OI} to {\OV}. The pLLS was first reported in the COS CGM Compendium (CCC) archival survey of \citet[CCC-I;][]{Lehner_2018}. The ionization parameter and metallicity ranges for the clouds are given in \citet[CCC-II;][]{Wotta_2019}, and \citet[CCC-III;][]{Lehner_2019}. We reanalyze the pLLS and the satellite cloud describing its full multi-phase properties, and also present new information on the galaxy environment. The paper is organized as follows. Sec. 2 gives details on the $HST$/COS archival spectra, and the $VLT$/MUSE, the $VLT$/VIMOS, and Magellan galaxy surveys. In Secs. 3 and 4, we present line measurements and the analysis of the chemical and physical state of the gas based on photoionization models. Information on galaxies is described in Sec. 5, with a discussion on scenarios that lead to an understanding of the origin of this absorber complex.  Throughout the paper, we adopt the cosmology with $H_0 = 69.6~\kms$~Mpc$^{-1}$, $\Omega_\textrm{m} = 0.286$ and $\Omega_{\Lambda} = 0.714$ from \citep{bennett_2014}. Projected separations are all in proper distance units. For solar elemental abundances, we adopt \citet{2009ARA&A..47..481A}, and \citet{grevesse_chemical_2010}. All the logarithmic values mentioned are in base-10.

\section{Spectroscopic observations}

The absorption line analysis is based on $HST$/COS observations of the QSO HE~$1003+0149$ ($z_{\mathrm{em}} = 1.080$) obtained under the Program ID. 12264 (PI. Simon Morris). The COS observations at intermediate resolution ($R \approx 20,000$) span the wavelength range $1140 - 1770$~{\AA} with total exposure times of $11.2$~ksec and $22.4$~ksec in the G130M and G160M far-UV gratings, respectively. The coadded spectrum was taken from the HST Spectroscopic Legacy Archive\footnote{https://archive.stsci.edu/hst/spectral\_legacy/} \citep{peepleshsla}. The spectra were resampled to two wavelength pixels per resolution element of $\Delta \lambda = 0.06$~{\AA}. The rebinned spectrum was continuum normalized using lower-order polynomials to define the continuum level. The $VLT$/UVES optical spectrum for this QSO, taken from the UVES Spectral Quasar Absorption Database (SQUAD) DR1 \citep{murphy_squad} of the ESO archive, has a low signal-to-noise ratio of $S/N \lesssim 10$~per 0.1~{\AA} resolution element across most parts of the spectrum. Nonetheless, we include it in our analysis as the optical spectrum covers a few important lower ionization lines including {\MgI},  {\MgIIdblt}, and {\FeII} multiplet transitions associated with the absorber.

The column densities, Doppler $b$-parameters and velocity centroids of the different absorption lines were obtained through best-fit Voigt profiles. The fitting was done using the VPFIT routine (ver 10.4)\footnote{https://people.ast.cam.ac.uk/~rfc/vpfit.html} with the model profiles convolved with the COS line-spread functions\footnote{https://www.stsci.edu/hst/instrumentation/cos/performance/spectral-resolution} for the nearest corresponding wavelength in the observed spectrum, and rebinned to the $\approx 0.06$~{\AA} spectral pixel width. Multiple lines from the same ion were fit together to constrain the column densities and $b$-parameters self-consistently. The initial guess for the line parameters in the VPFIT procedure for ions that are potentially tracing gas of similar ionization (such {\CII} and  {\OII}, {\CIII} and {\OIII}) were guided by each other. The column densities were also measured by integrating the pixel-by-pixel apparent optical depth (AOD) across the absorption feature following the technique given by \citet{savage_analysis_1991}. The comparison of column densities from AOD measurements and VP fits helps to identify line saturation unresolved by COS. For lines which are non-detections, an upper limit on the column density was arrived at from the $3\sigma$ upper limit on the equivalent width, assuming the linear part of the curve-of-growth. 

\section{Galaxy data}

The HE~$1003+0149$ quasar field was observed using MUSE \citep{BaconMUSE} under PID: 095.A-0200(A) (PI: Schaye) on 2015-04-19 for a total exposure time of 2h (on source) with wide-field mode (field-of-view (FoV) of $1^{\prime}~\times~1^{\prime}$), with nominal spectral coverage ($4750-9300$~\AA), and without adaptive optics. Each observing block of 1h was divided into $4\times900$ s exposures. Frames were rotated by 90 degree and offset by a few pixels to minimize the systematic uncertainties. The 8 raw frames were reduced using the standard MUSE data reduction pipeline \citep[v1.2;][]{Weilbacher20} with default sets of parameters. The reduced frames were subsequently post-processed using the {\sc CubeFix} and {\sc CubeSharp} tools in the CubExtractor package \citep[]{Cantalupo19} to improve the flat-fielding and sky-subtraction respectively, following the steps detailed in \citet{Marino18}. The effective seeing of the final coadded MUSE cube is  $0.9^{\prime\prime}$  at $\approx 7000$~\AA. The FoV of MUSE centered on the quasar allows us to probe galaxies up to an impact parameter of $\approx 230$~kpc at the redshift of the absorber. The spectral resolution of MUSE is $\approx 120$~\kms at the expected position of the [\OII] emission from galaxies at the redshift of the absorber. 

We run {\sc Source Extractor} \citep[SExtractor][]{sextractor}  on the white-light image constructed from the MUSE cube with a detection threshold of $1\sigma$ per pixel ($\rm DETECT\_TRESH = 1$) and requiring a minimum number of neighbouring pixels above the
threshold of 3 ($\rm DETECT\_MINAREA = 3$). A modified version of the application {\sc Marz} \citep{hinton16} was then used to classify the SExtractor-detected objects and determine their redshifts. The redshifts were further refined using a modified version of the code {\sc Platefit} \citep{Brinchmann04} by fitting Gaussian profiles to the available emission and absorption line features. 

We used the [\OII] line fluxes returned by {\sc Platefit} for calculating the SFR using the calibration relation of \citet{Kewley04} adjusted for the \citet{Chabrier03} initial mass function (IMF). The stellar masses of the galaxies were estimated using the stellar population synthesis (SPS) code {\sc FAST} \citep[v1.0]{Kriek09} by creating 11 pseudo-narrow bands (400~\AA) constructed out of the MUSE spectrum after masking the emission line features. The halo masses are subsequently estimated using the abundance matching relation from \citet{Moster13}. The details of the identification procedure and determinination of galaxy properties will be presented elsewhere. In total we have identified 3 galaxies within the MUSE FoV and within $\Delta v = \pm 300$~\kms of the absorber. 
     
Additional galaxy information within 5~Mpc and $|\Delta v| \leq 1000$~{\kms} of the projected field around the absorber was obtained from the VIMOS VLT Deep Survey \citep[VVDS\footnote{https://cesam.lam.fr/vvds/}][]{fevre2013vimos}, and the Magellan galaxy redshift survey. VVDS is a galaxy spectroscopic redshift survey  using the VIMOS multi-slit spectrograph at VLT. The data set is a composite of three different magnitude limited surveys. The widest sky coverage among these has galaxies in the magnitude range of $17.5 \leq i \leq 22.5$. The other two surveys go deeper in magnitude but narrower in sky coverage. The wide field survey dominates the VVDS sample. A $100$\% completeness for the composite sample down to an approximate $i$-band magnitude of $m \approx 23$, converts into a luminosity threshold of $L \gtrsim 0.2L^*$ at $z \approx 0.8$ using the Schechter luminosity function parameters given in \citet{2005ApJ...631..126D}. Information on the VVDS galaxies for this field is also given in the {\HI}-galaxy cross-correlation catalogue of \citet{10.1093/mnras/stt1844}.  We have also extracted an archival $HST$/ACS wide-field image centered on the quasar field (Prop ID: 14269, PI: Nicholas Lehner). The single band image obtained with the F814W filter over the band-pass of $7000 - 9750$~{\AA} has an exposure time of $2.18$~ks. 

The Magellan galaxy redshift survey data were acquired as part of a deep and highly complete galaxy redshift survey in the fields of COS quasars as described in \citet{Chen2009}, and \citet{Johnson_2015}, and briefly summarized here. The absorption-blind survey targeted galaxies as faint as $r = 23$ (AB system) using the IMACS \citep{Dressler2011}, and LDSS3 spectrographs in the multi-slit mode. The IMACS spectra were acquired with the f2 camera, and the 200l grism, and the LDSS3 spectra were acquired with the VPH-all grism. The multi-slit data were reduced with the COSMOS pipeline \citep[see][]{Dressler2011,Oemler2017}, and redshifts measured using the SDSS BOSS galaxy eigenspectra from \citet{Bolton2012}.

\section{Spectroscopic Analysis of the Absorption System}

\begin{figure*} 
   \centering \includegraphics[scale=1.0]{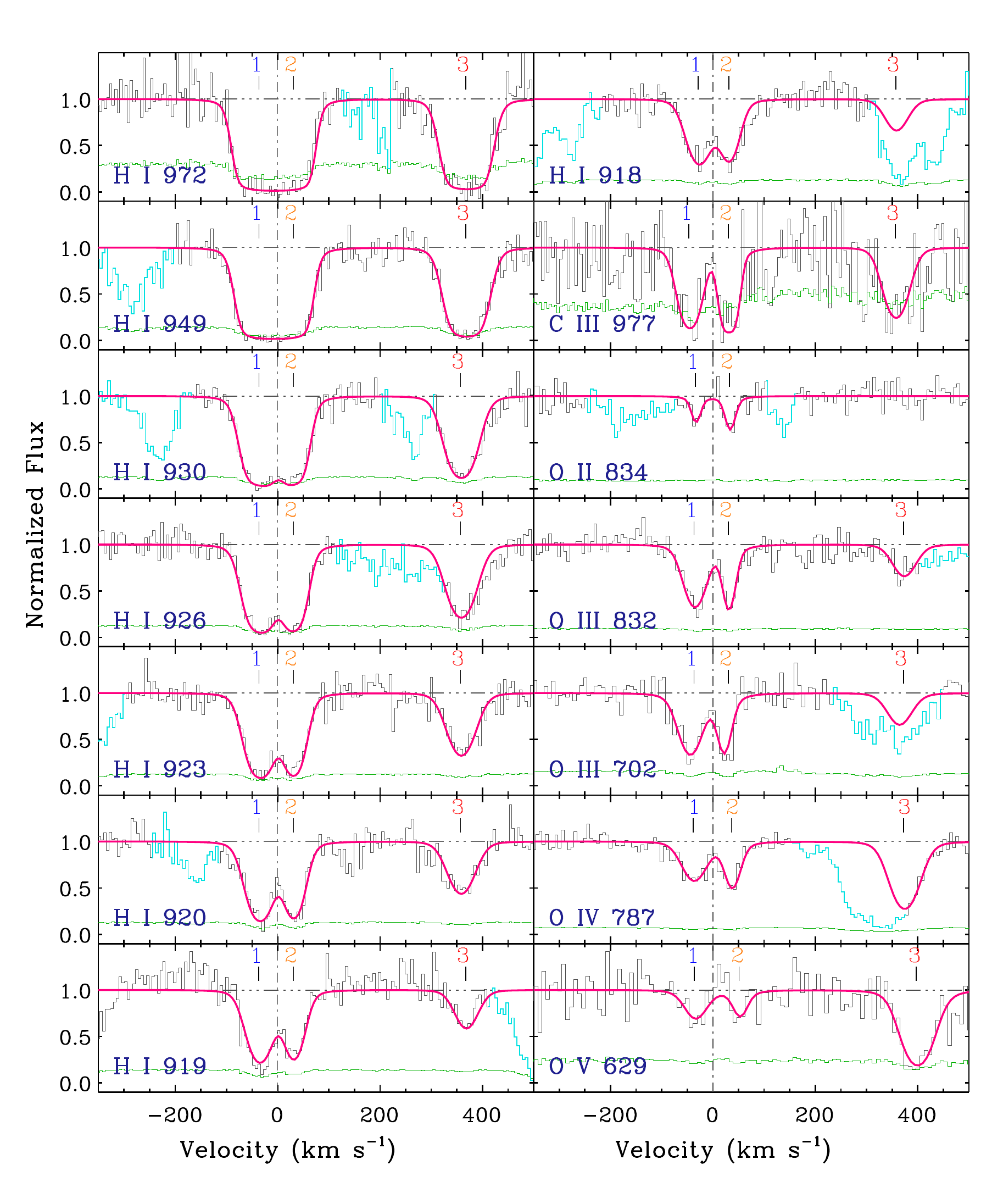}
   \caption{The hydrogen and metal-lines associated with the $z = 0.83718$ absorber complex. Each panel shows the segment of the COS spectra featuring the relevant absorption line. The X-axis is velocity in the rest-frame of the absorber. The three kinematically distinct components are labeled as 1, 2, and 3, with the Voigt profile models superimposed. The $1\sigma$ error spectrum is plotted at the bottom of each panel in \textit{green}. Absorption due to component 3 from {\HI}~$918$ and higher orders overlaps with the higher order Lyman transitions from the central two clouds. The higher order Lyman transitions are included in Figure~\ref{fig:pLLS}. Features unrelated to the system are shown in \textit{cyan} color in the individual panels.}
    \label{fig:sysplot}
\end{figure*}

\begin{figure*}
   \includegraphics[scale=0.8,angle=180]{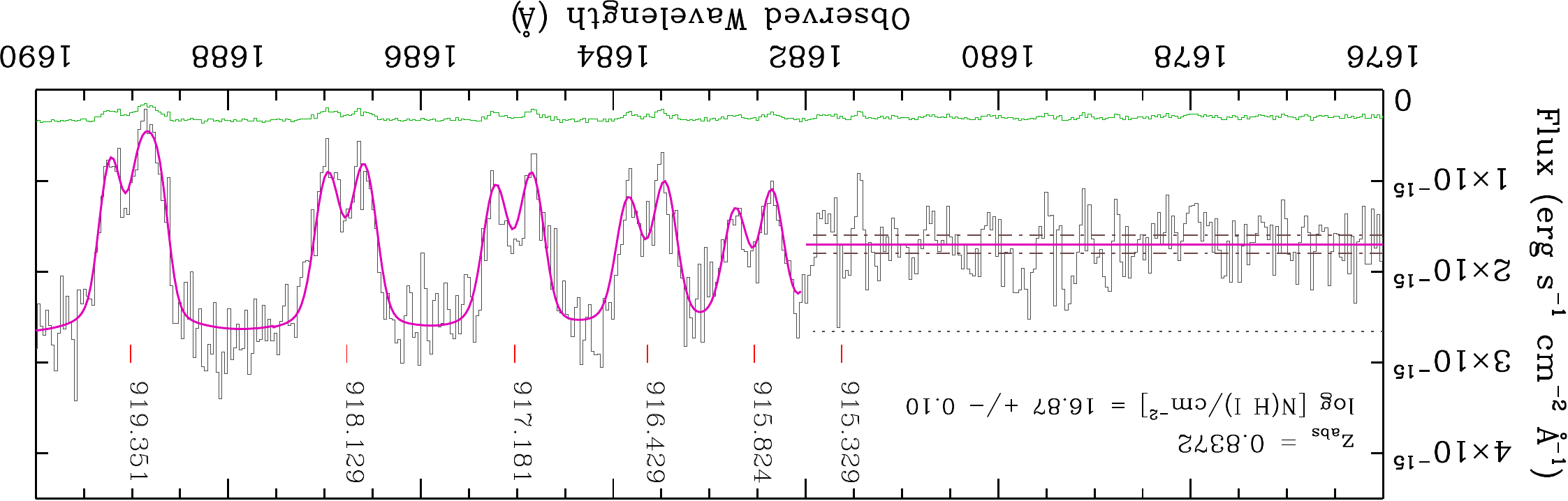}
   \caption{A segment of the COS spectrum of HE~$1003+0149$ with the partial Lyman limit break due to the $z = 0.83718$ absorber. The higher order Lyman lines due to the central components (components 1 and 2 of Figure  \ref{fig:sysplot}) are labeled, and the Voigt profiles are superimposed. The {\textit{dash-dot}} line indicates the two extreme continuum levels adopted to estimate the uncertainty in the {\HI} column density. The optical depth at the Lyman-limit is $\tau_{912} = 0.399$, and the corresponding {\HI} column density is $\log~[N(\HI)/\cmsq] = 16.87$.}
    \label{fig:pLLS}
\end{figure*}

Figure \ref{fig:sysplot} shows the {\HI} and metal lines in the absorber's rest frame. The line measurements are listed in Table \ref{table1}. The absorber has three distinct kinematic components, two of them close-by with a line-of-sight separation of $|\Delta v| \approx 70$~{\kms}, and the third one at $|\Delta v| \approx 400$~{\kms} from the twin components. The strengths of the higher order Lyman transitions and the spectrum around the partial Lyman limit break shown in Figure \ref{fig:pLLS} clearly indicate that the {\HI} column densities of the twin components exceed that of the component isolated in velocity at $\approx 400$~{\kms}. We have arbitrarily chosen the wavelength pixel between the twin components to define the redshift of the absorber complex as $z = 0.83718$ corresponding to $v = 0$~{\kms}. We proceed by referring to these central components as components 1 and 2, and the offset component isolated in velocity as component 3, as labeled in Figure \ref{fig:sysplot}.

The {\HI} in this absorber complex is detected in a range of higher order Lyman series lines from {\HI}~$972$ up to the Lyman limit. The {\Lya} and {\Lyb} are outside of the wavelength coverage of the COS G160M grating. Simultaneous Voigt profile fits to the Lyman transitions constrain the {\HI} column densities in the three separate components to values in the range of $\log [N(\HI)/\cmsq] =  16.14 - 16.56$. The {\HI} absorption in the offset component 3 starts to overlap with the absorption in the central twin components from $\HI~918$~{\AA} and higher orders at bluer wavelengths. The contaminating influence due to this overlap on the central components is only minor, as the offset cloud is much weaker at higher Lyman orders. The {\HI} column along the sightline, largely contributed by components 1 and 2, result in a partial Lyman limit break at $1682$~{\AA} (see Figure \ref{fig:pLLS}). The $\log~[N(\HI)/{\cmsq}] = 16.87~{\pm}~0.10$ estimated from the optical depth at the partial Lyman limit break closely matches the cumulative {\HI} column density of $\log~[N(\HI)/{\cmsq}] = 16.88~{\pm}~0.15$ from the three separate components, and the cumulative $\log~[N(\HI)/{\cmsq}] = 16.80~{\pm}~0.13$ from components 1 and 2 to within their $1\sigma$ uncertainties. \citet{Lehner_2018} list $\log N(\HI)$ of $16.52,~16.36,$ and $16.13$ dex for the three components respectively, which are in agreement with our measurements. 

Coincident with the double component {\HI} at $v \approx 0$~{\kms} (components 1 and 2) are lines from {\OII}, {\OIII}, {\OIV}, {\OV}, and {\CIII} detected at $> 3\sigma$ significance. The {\CII}, {\SIII}, {\SIV}, {\SV}, {\SVI}, {\NeVIII} and {\MgX} ions are important non-detections. The non-detection of {\CII} and the detection of {\OII} suggest non-solar [C/O] abundances in components 1 and 2 since both ions are tracers of gas of similar ionization. The $> 3\sigma$ absorption detected at the expected location of {\SII}~$765$ was found to be inconsistent with being {\SII}, as explained in the Appendix. The {\CIII}~$977$~{\AA} is saturated and falls at the edge of the G160M grating where the $S/N$ is poor. To improve the constraints on the column density during VP fitting, the $b$-parameter of {\OIII} is adopted for {\CIII} as well. The creation and destruction energies of these two ions are similar ($24/48$~eV, and $35/55$~eV respectively), and therefore trace the same phase of gas. The component 1 line profiles are broader than component 2, with metal line $b$-values of $\approx 22$~{\kms} and $\approx 12$~{\kms} respectively. The $b$-parameter of {\HI} and metal lines measured at the resolution of COS are comparable in component 1 suggesting a predominance of non-thermal broadening. Even in component 2, the $b(\HI) = 21~{\pm}~3$~{\kms} and $b(\OIII) = 10~{\pm}~4$~{\kms} yield a temperature of $T = 2.2_{-1.4}^{+1.3} \times 10^4$~K with non-thermal mechanisms dominating the {\HI} line broadening. The temperature range here is based on simultaneously considering the $1\sigma$ limits on $b(\HI)$ and $b(\OIII)$.   

Coinciding with the offset {\HI} (component 3) are lines from {\CIII}, {\OIII}, {\OIV}, and {\OV}. The {\CII}, {\OII}, {\SII}, {\SIV}, {\SV}, {\SVI}, {\NeVIII}, and {\MgX} ions are key non-detections. The {\OV}~$629$~{\AA} in this component is a saturated feature. The $\approx 0.07$~dex difference between the AOD integrated column density and profile fit value for this line indicates that the saturation is only mild, with the profile fit potentially recovering the true column density. The {\OIV}~$787$~{\AA} line suffers severe contamination from {\OIII}~$702$~{\AA} associated with the quasar itself. The {\OIV}~$787$~{\AA} profile was generated by adopting the same line centroid as the {\OIII}~$832$~{\AA} line, and by deweighting the pixels from $+160$~{\kms} to $+360$~{\kms}, assuming that the red wing of the observed profile is contributed by {\OIV}. The uncertainty of $0.08$~dex for the {\OIV} column density obtained from the profile fitting is a significant underestimation as it does not take into account the ambiguity due to contamination. We thus adopt the measured $\log~[N(\OIV)/{\cmsq}] = 14.6$ as a conservative upper limit on the column density.

The {\OIII}~$702,~832$~{\AA} lines in this component also suffer from different levels of contamination. The {\OIII}~$832$~{\AA} line, where the contamination appears to be minimal, is fitted simultaneously with {\CIII} to better constrain the line parameters. The corresponding fit parameters for {\OIII} is used to synthesize the {\OIII}~$702$~{\AA} which is then superposed at the expected location to discern the extent of contamination (refer to Figure \ref{fig:sysplot}). The $b$-values of {\HI} and metal lines are comparable, suggesting a predominance of non-thermal line broadening.  

The relative strengths of the oxygen ions between the three clouds allude to ionization differences along the line of sight.  In component 3, we measure $N(\OV) > N(\OIII) > N(\OII)$, whereas in components 1 and 2, {\OIII} is stronger compared to {\OIV} and {\OV}. Furthermore, {\OII} is a formal detection in the central two components, whereas it is a non-detection in component 3. This suggests lower ionization conditions in the central components compared to the offset component. 

The temperatures of $T \approx (1.2 - 2.3) \times 10^4$~K indicated by the respective line widths for the three components favor photoionization as the dominant mechanism regulating the ionization. The information on five successive stages of oxygen from {\OII} to {\OV} spanning ionization energies in the range of $35 - 138$~eV can provide the constraints needed for establishing the physical conditions in the absorbing gas. An upper limit on column density is also available for {\OI} from the undetected {\OI}~$971$~{\AA} line. Here we note that many Lyman limit and partial Lyman limit systems are multiphased. The presence of separate low-and high-ionization phases often becomes evident only through velocity offsets, and/or velocity widths between the low ions ({\OII}, {\CII}, {\SiII}, {\CIII}, {\OIII}) and the high ions, particularly {\OVI} \citep{Lehner_2013,crighton2013metal,2013ApJ...778..187F,fumagalli2011absorption}. For the three distinct clouds in the absorber, noticeable differences in line widths or line-of-sight velocities are not seen between the ions. Nonetheless, the ionization modeling approach that we adopt reveals the presence of more than one gas phase in the absorbing clouds. 

\section{Physical properties \& chemical abundances of the absorber}

To determine the ionization conditions and the chemical abundances in the absorber, we implement Cloudy \citep[ver C17.01][]{2013RMxAA..49..137F} photoionization equilibrium (PIE) models. These models assume the absorbing medium to be of uniform density with a constant temperature and plane-parallel geometry. We adopt the solar relative elemental abundance pattern given by \citet{2009ARA&A..47..481A} in our  models. We further assume that the photoionization in the cloud is regulated by the extragalactic background radiation (EBR) at the absorber redshift as given by \citet[][hereafter KS19]{khaire2019new}. The ionizing background is likely to be dominated by this EBR, since the nearest detected galaxies are dwarfs, sufficiently distant from the absorber with projected separations of $\rho > 150$~kpc, and $\rho/R_{\mathrm{vir}} > 1$ (see Table~\ref{tab:Table 2a}). The three kinematically distinct components are modeled separately. 

We adopt the Bayesian approach described in \citet{Sameer_2021} for the component-wise multiphase modeling of the absorption system. The {\OII} was used as the optimizing ion for the low ionization phase of components 1 and 2. This means that the total column density of \textsc{cloudy} models was adjusted until the models converged on the measured column density of the optimized transition. Since {\OII} is a non-detection for component 3, {\OIII} was chosen as the optimizing ion for that component. If the observed transitions could not be simultaneously fit with this low ionization phase, then separate high ionization clouds were considered, optimized on {\OV}. A grid of Cloudy models were generated for metallicities (relative to solar) and ionization parameters in the range of $\log~(Z/Z_{\odot})$ = [ $-4,1.5$], and $\log~U = [-5.0,0.0]$\footnote{Ionization parameter is the ratio of number density of hydrogen ionizing photons to hydrogen number density, $\log~U = \log n_{\gamma} - \log n_{\H}$. It thus serves as a proxy for density.} respectively with a step size of 0.1 dex. The models are further interpolated to a step size of 0.01 dex. The models are allowed to run until convergence such that the observed column density, determined from the profile fitting of the optimized ion, is recovered for that component. For this convergence solution, the column densities of the other ions are extracted from the model output. Using the photoionization equilibrium temperature predicted by the model, we estimate the Doppler broadening parameter $b$ for all the other transitions using the expression $b^{2}=b^{2}_{\mathrm{nt}}+b^{2}_{\mathrm{t}}$, where $b^{2}_{\mathrm{t}}=2kT/m$ is the line broadening solely due to temperature and $b^{2}_{\mathrm{nt}}$ the line broadening due to turbulence. This non-thermal component to line broadening is determined from the optimized ion with the knowledge of temperature from the Cloudy model. The non-thermal broadening component is assumed to be identical for other transitions in the same phase. 

\begin{figure*}
   \includegraphics[scale=0.5]{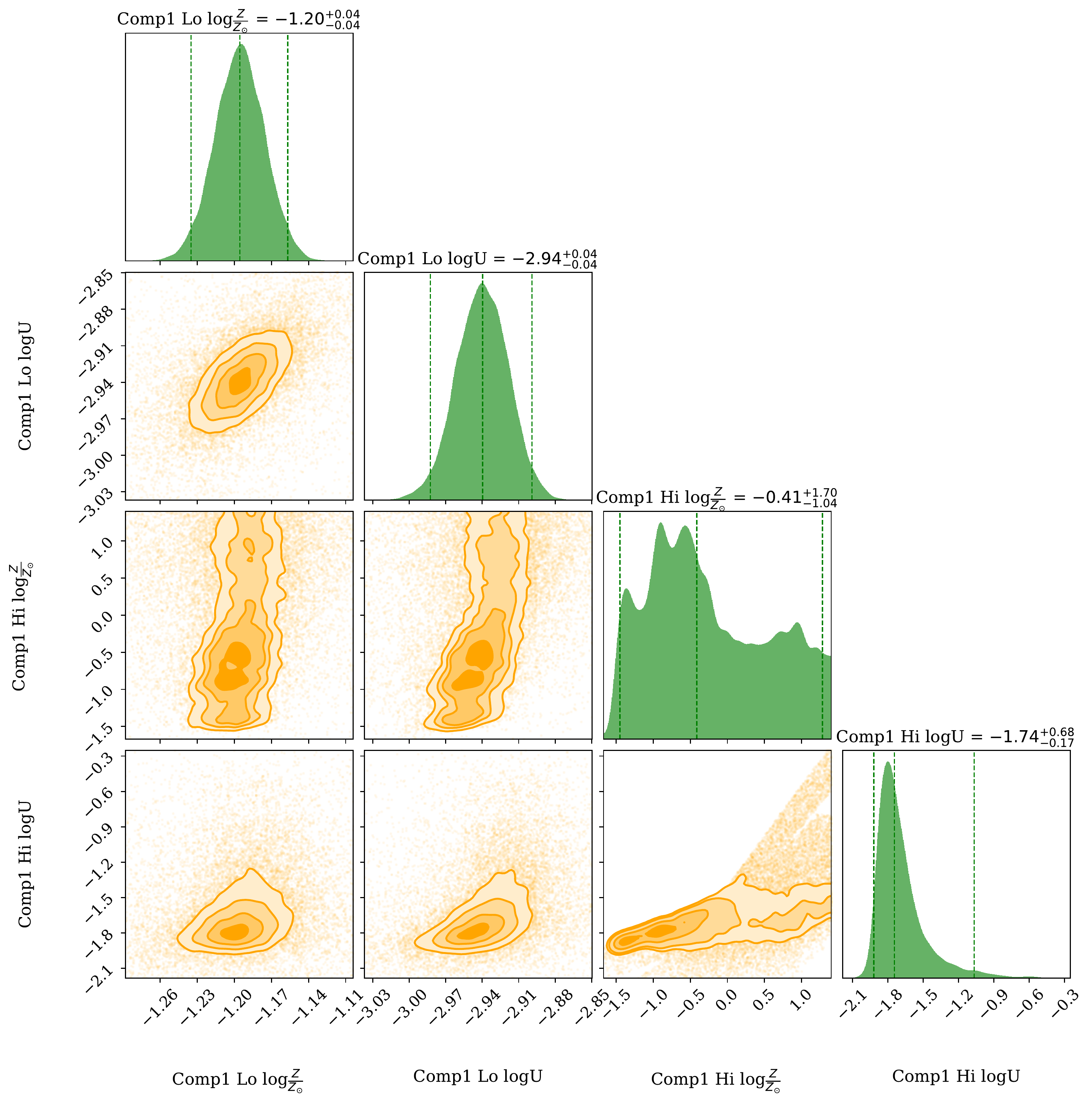}
   \caption{The corner plot showing the marginalized posterior distributions for the metallicity ($\log~(Z/Z_{\odot})$) and ionization parameter ($\log~U$) for component 1 in the low phase (traced by {\OII}) and high phase (traced by {\OV}). The low ionization phase is indicated by "Lo", and the high ionization phase by "Hi". The over-plotted vertical lines in the posterior distributions span the 95\% credible interval. The contours indicate 0.5$\sigma$, 1$\sigma$, 1.5$\sigma$, and 2$\sigma$ confidence levels. The model results are summarised in 
   Table \ref{tab:Table 3}, and the synthetic profiles based on these models are shown in Figure~\ref{fig:systemplotmodel2}.}
    \label{fig:corner1}
\end{figure*}

\begin{figure*}
   \includegraphics[scale=0.5]{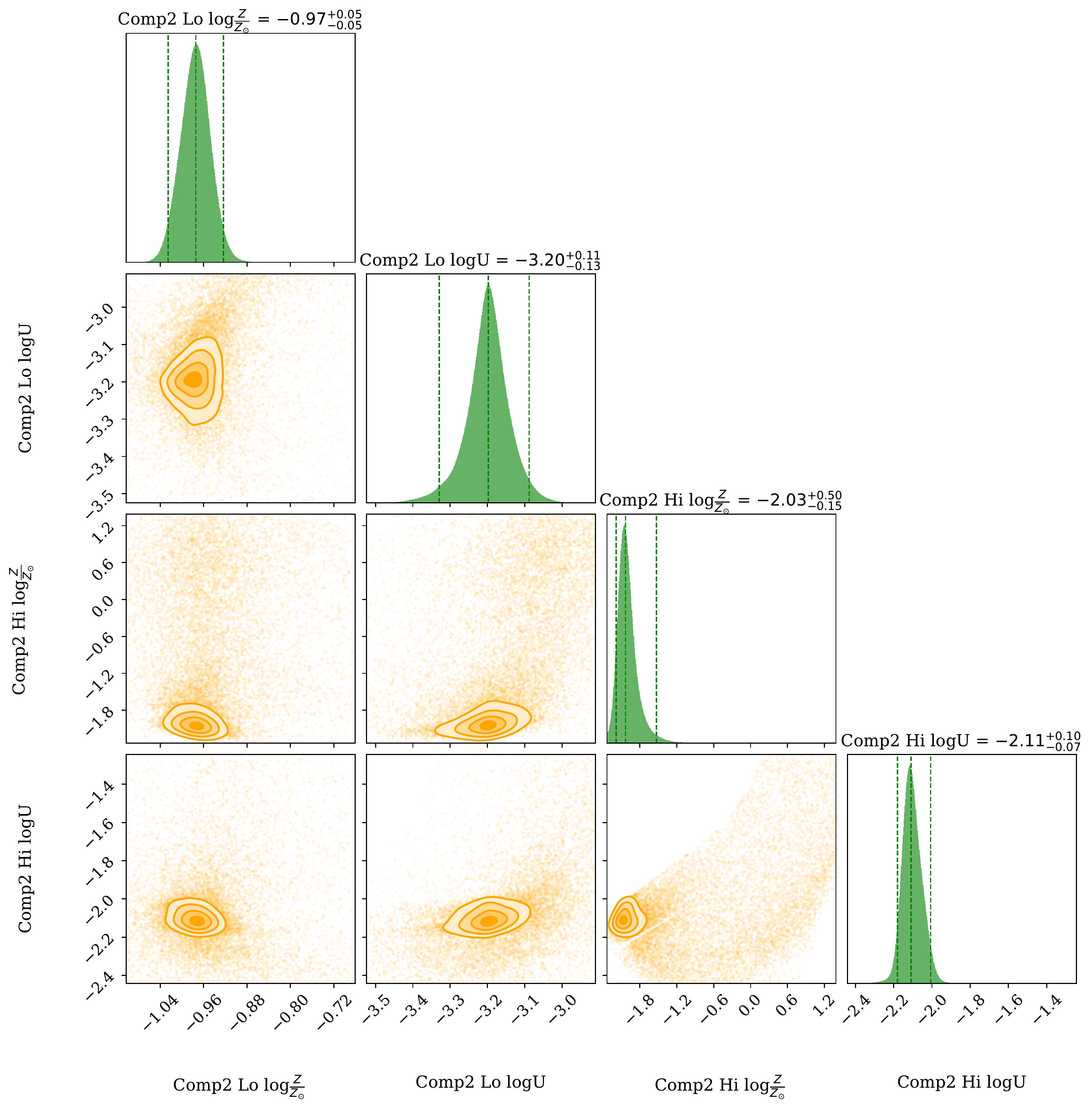}
   \caption{As Figure~\ref{fig:corner1}, but for component 2.}
    \label{fig:corner2}
\end{figure*}

\begin{figure*}
   \includegraphics[scale=0.5]{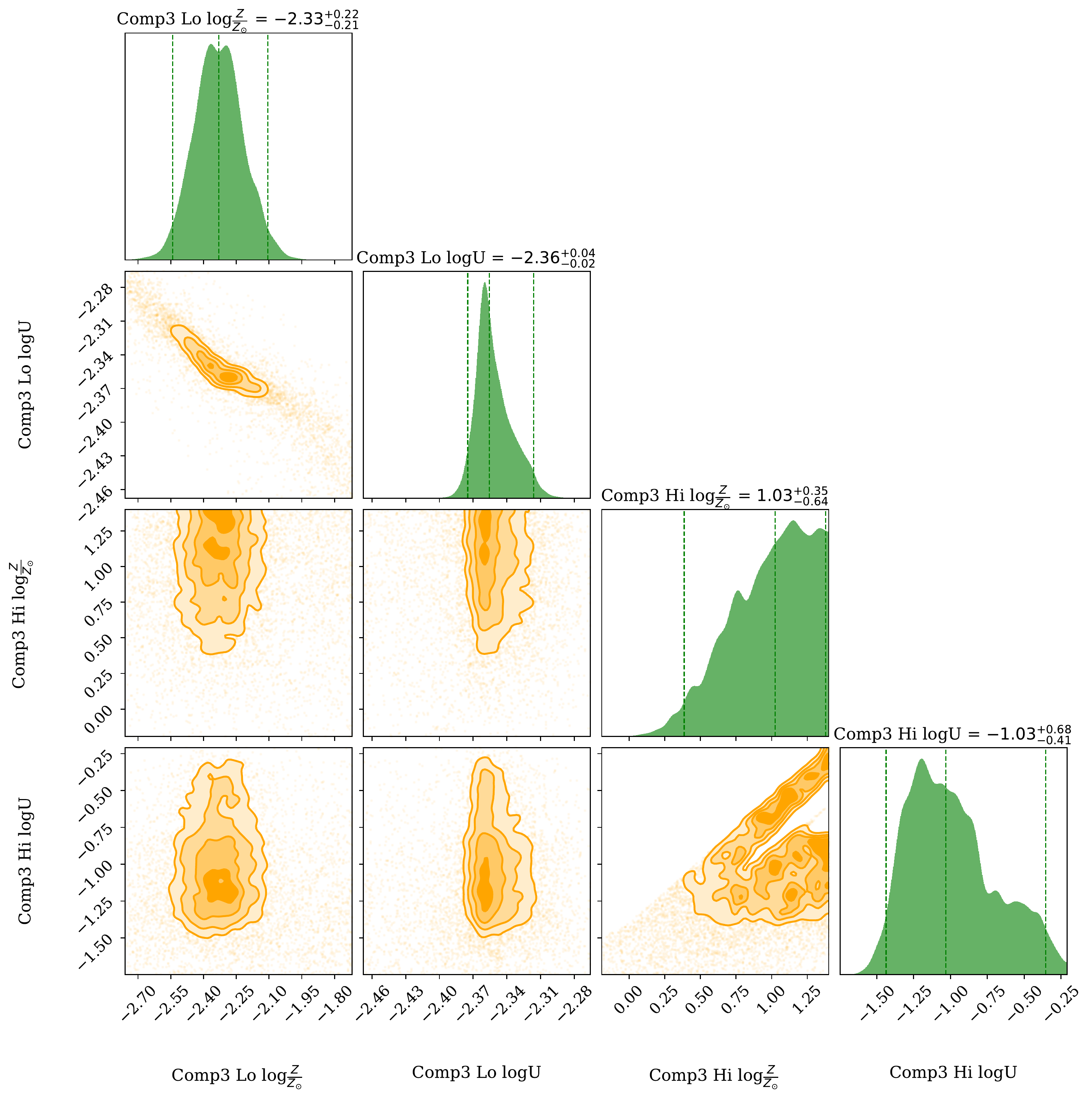}
   \caption{As Figure~\ref{fig:corner1}, but for component 3.}
    \label{fig:corner3}
\end{figure*}

\begin{figure*}
   \includegraphics[scale=0.4]{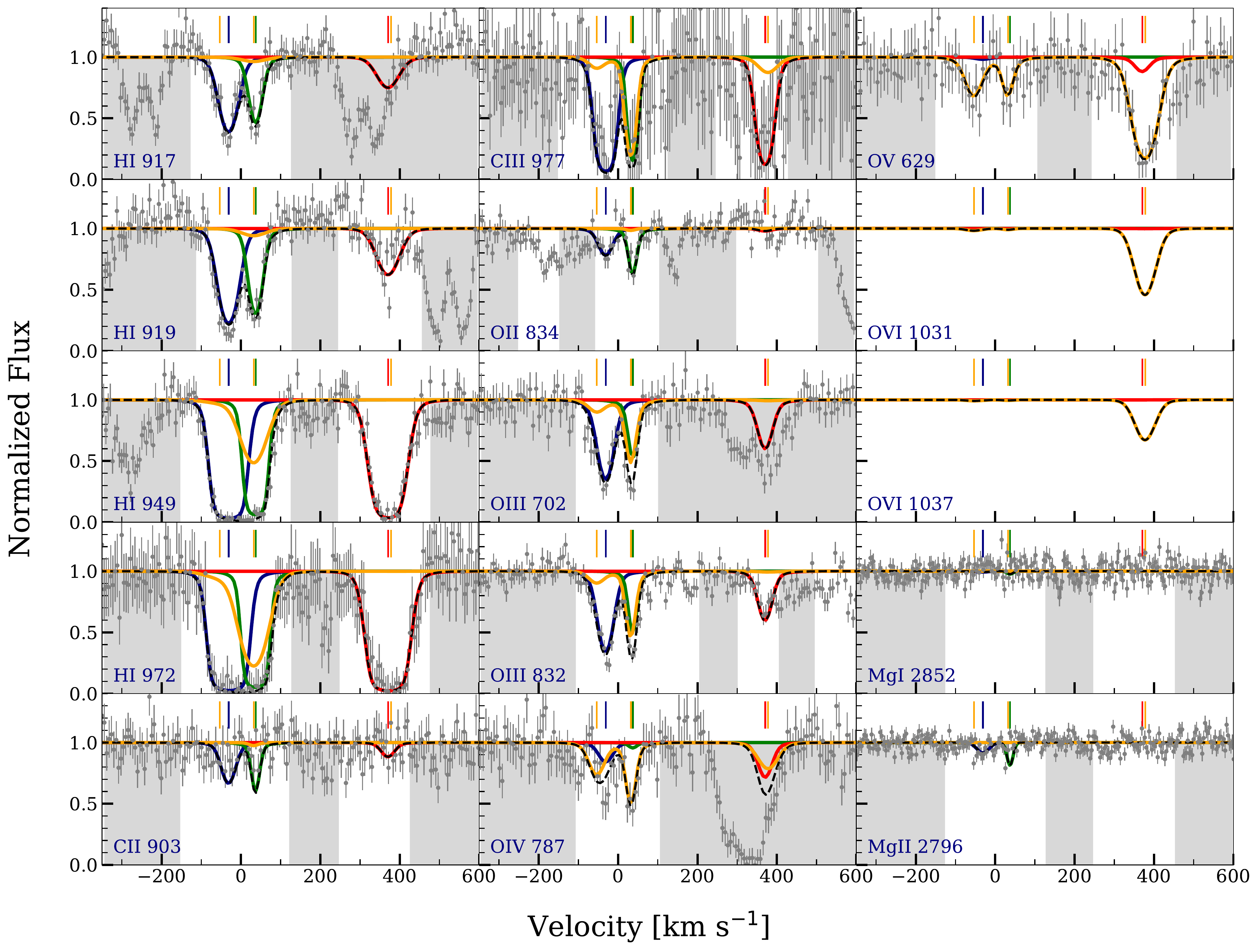}
   \caption{The synthetic profiles based on the maximum likelihood estimate values of the ionization models of Figures 3, 4 and 5, and summarized in Table \ref{tab:Table 3}, are shown overlaid on the data. The absorption due to the low ionization gas phases of components 1, 2, and 3 are depicted using \textit{blue, green,} and \textit{red} synthetic profiles respectively. The {\HI} absorption is entirely coming from the low ionization phase in the three clouds. As the models show, the {\OV} in all the three components require a separate phase of higher ionization gas which is indicated by the \textit{orange} profiles. This high ionization gas also contributes partially to the {\OIV} absorption. The expected absorption from {\OVIdblt} lines for the three components are also synthesized and shown, though the available spectra do not cover this transition. In moderate $S/N$ spectra, similar to COS, only {\OVI} associated with component 3 is likely to be a formal $\geq 3\sigma$ detection. The {\CII} components are overproduced by the models, suggesting [C/O] $< 0$. The {\OIII}~$702$~{\AA} and {\OIV}~$787$~{\AA} lines of component 3 are contaminated. The shaded regions are masked during the log-likelihood evaluation.}
    \label{fig:systemplotmodel2}
\end{figure*}

With the estimates thus arrived at for $b$, $N$, and $v$ for each ion, the Voigt profiles for all transitions are synthesized using the appropriate line spread function for the instrument and over-plotted on the observed lines. Only the observed VP fit column density of the optimized transition is used in our analysis. The method of superimposing synthetic spectra can better constrain the model parameters because it utilizes the shapes of the observed profiles, and considers how multiple phases combine to produce intermediate ionization transitions. This is particularly appropriate for the {\HI}~series lines where the precise shapes of the profiles on the flat part of the curve of growth, and self-consistency with weaker lines, can constrain metallicity. Such a comparison helps to discern whether the ionic species detected in the three individual components are from the same phase and which ones require a separate phase. 

For metallicity [$\log~(Z/Z_{\odot})$], and ionization parameter ($\log~U$) estimations, in case of components 1 and 2 we make use of the Lyman series transitions, {\OII} $\lambda$834, {\OIII} $\lambda$832, {\OIII} $\lambda$702, {\OIV} $\lambda$787, {\OV} $\lambda$629, {\CIII} $\lambda$977 from the COS G130/G160M spectrum, and make use of {\MgII} $\lambda$2796, {\MgII} $\lambda$2803, {\MgI}~$2853$, {\FeII} $\lambda$2382, {\FeII} $\lambda$2600, and {\FeII} $\lambda$2586 from the UVES spectrum. However, in case of component 3, because of blending issues we ignore the {\HI} $\lambda$917, {\HI} $\lambda$918, {\OIII} $\lambda$702, {\OIV} $\lambda$787 transitions while including all the other transitions mentioned earlier. 
The corner plots in Figures 3, 4 and 5 show the posterior distributions for metallicity and density for components 1, 2, and 3 respectively, and the modeling results are summarized in Table \ref{tab:Table 3}. The synthetic profiles based on these models, convolved with the relevant instrumental spread function, superposed on the data are shown in Figure \ref{fig:systemplotmodel2}. The column density estimates determined from these models agree well with the measured values within uncertainties in all the cases, except for the {\OIV} measurement in component 3 as it is severely affected by blending, preventing a secure measurement using VPFIT. The absorption in all three components require a two-phase solution, with the {\OV} tracing higher-ionization gas compared to {\OII} and {\OIII}. In component 1, the observed {\OII} column density requires [O/H] $\geq -1.5$. For oxygen abundance lower than this, the PIE models overproduce {\HI} at all densities. As the analysis of Figure \ref{fig:corner1} shows, the ionization solution that simultaneously explains {\OII} and {\OIII} in this cloud has a density of $n_{\H} = (6.3~\pm~0.6) \times 10^{-3}$~{\cc} corresponding to an ionization parameter range of $\log U = (-2.94~\pm~0.04)$, and [O/H] = $-1.20~\pm~0.04$. The listed uncertainty corresponds to the 95\% credible interval ($2\sigma$) from the posterior distribution. There can be an uncertainty of $\approx 0.2$~dex due to the choice of the EBR model. In the ionization models we have used the fiducial Q18 model of KS19, which is one of a range of observationaly consistent quasar SEDs in the KS19 ionizing background radiation models (Acharya \& Khaire in prep.) Such a phase also explains the observed {\CIII} for [C/O] = 0. However, we caution that the relative abundance estimate can be significantly affected by the strong saturation of the {\CIII}~$977$ line. Interestingly, the {\CII} predicted from this low-ionization phase assuming solar abundance exceeds the upper limit set by the {\CII}~$903.6$ and {\CII}~$903.9$~{\AA} lines, implying that the relative abundance of [C/O] is sub-solar. The {\HI} given by the model for the maximum likelihood metallicity matches well with the observed {\HI} in this cloud, as expected from the common scenario of hydrogen predominantly tracing the lowest ionization gas in multiphase absorbers. 

A similar low-ionization phase is also identified by the models for component 2 with $n_{\H} = 1.2^{+0.4}_{-0.3} \times 10^{-2}$~{\cc}, and [O/H] = $-0.97~\pm~0.05$, with the hydrogen coincident in velocity with component 2 predominantly tracing this phase. Here again, the predicted {\CII} exceeds the upper limit, implying [C/O] $< 0$. To estimate the [C/O], we generate photoionization models for different [C/O] values ranging between [-1.5, 0.0] in intervals of 0.05 dex corresponding to the maximum likelihood values of metallicity and ionization parameter. We again use the Bayesian sampler to explore this grid and find the models that explain the observed data. Allowing for variation in [C/O], we determine [C/O] = $-0.25^{+0.21}_{-0.69}$, and $-0.24^{+0.21}_{-0.82}$ for components 1 and 2, respectively, where the uncertainties are $2\sigma$ limits. 

Interestingly, the low-ionization gas in components 1 and 2 significantly under-produces {\OV}, and also does not fully explain {\OIV}, implying the presence of multiphase gas. The properties of a separate higher-ionization phase were determined by optimizing the models on {\OV}. The lack of additional constraints for this phase (such as information on {\OVI}) resulted in wider posterior distributions and less constrained estimates for the ionization parameter and metallicity (see the summary of modeling results in Table \ref{tab:Table 3}). For component 3 also, a two-phase solution consisting of a low-ionization phase traced by {\OIII} and a high-ionization phase traced by {\OV} was found necessary to explain both ions. The low-ionization phase produces little {\OII} consistent with its non-detection in the COS spectrum.  The amount of {\HI} arising from the high-ionization phase is low and can vary by an order of magnitude preventing us from constraining its metallicity. Even with the uncertainty, the models for all three clouds predict an {\HI} column density in the high-ionization gas that is $\approx 1 - 3$~dex lower than the corresponding low ionization phase. For the high-ionization phase in components 1 and 3, we place lower limits on the metallicity as a broad range of values are deemed feasible as shown by the posteriors in Figure~\ref{fig:corner1} \& \ref{fig:corner3}. Differences in metal abundances of an order of magnitude or more between kinematically coincident phases, and closely separated components can result from inadequate mixing of metals at small scales \citep[e.g.,][]{Rosenwasser2018,Zahedy2019,Sankar2020}, with the presence of a magnetic field enhancing the scatter in metallicity further \citep{vandevoort2021}. Our modeling approach reveals how kinematically coincident components can trace gas of different density, temperature, and metallicity. 

\citet{Wotta_2019} have carried out analysis of components 1 and 2. Component 3 is classified in their scheme as a high column density {\Lya} forest absorber, and hence excluded from their analysis which was exclusive to pLLS and LLS. Instead, in a follow-up work on strong {\Lya} absorbers, the ionization results for component 3 were included \citep{Lehner_2019}. The density ranges arrived at by \citet{Wotta_2019} are comparable to the results we obtain. The metallicities in their analysis are derived using the low ions ($\OI$, $\OII$) as the primary constraints under the assumption that species such as {\OIII}, {\OIV} and {\OV} are prone to arise in separate phases. For components 1 and 2 they determine $\log~(Z/Z_{\odot})= -1.5~\pm~0.1$ and $\log~(Z/Z_{\odot}) = -2.2~\pm~0.2$, respectively (the latter value from Lehner et al. 2019). Their estimate for component 2 is $\sim 1.2$~dex lower than what we arrive at\footnote{For component 2, \citet{Wotta_2019} quote a metallicity of $\log~(Z/Z_{\odot}) = -1.9$, but in \citet{Lehner_2019} the value is updated to $-2.2$. We have compared against this updated value.}. We note that while modeling they have adopted an upper limit of $\log~[N(\OII)/\cmsq] \leq 13.21$ for component 2 and a measurement of $\log~[N(\OII)/\cmsq] = 13.42~\pm~0.16$ for component 1, though in the data, the {\OII} absorption in component 2 is stronger than in component 1. The higher column of {\OII} coupled with the corresponding lower column on {\HI} in component 2 should result in a higher value for [O/H] compared to component 1, as we find in our analysis. For component 3, \citet{Lehner_2019} estimate an $80$\% confidence upper limit of $\log~(Z/Z_{\odot}) < - 1.74$, with a [C/$\alpha$] = $-0.06^{+0.29}_{-0.40}$. The metallicity upper limit is based exclusively on the non-detections of the low ions. We estimate an [O/H] $= -2.33^{+0.22}_{-0.21}$ based on the {\OIII}, which is consistent with the \citet{Lehner_2019} upper limit, and also reproduces the observed {\HI} in this cloud from the same phase. Additionally, our models also suggest a higher ionization phase in this cloud to explain the {\OV}. The predicted column density of {\OVI} from this high phase is $\log [N(\OVI)/\cmsq] =$ 13.9 $\pm$ 0.8. The predicted {\OVI} corresponding to the maximum likelihood estimate parameters is shown in Figure~\ref{fig:systemplotmodel2}.

\section{Galaxies near the absorber \& the possible origins of the absorber complex}
\label{sec:discussion}
The quasar field is covered by the MUSEQuBES  (MUSE Quasar-field Blind Emitters Survey), which is a blind survey for {\Lya} emitters in the $1^{\prime} \times 1^{\prime}$ fields around bright quasars, and the VIMOS VLT Deep Survey (VVDS)\footnote{http://cesam.lam.fr/vvds/} galaxy multi-object spectroscopic surveys. The MUSE data reduction is described in \citet{2020MNRAS.496.1013M}. The current data release of VVDS is given in \citet{fevre2013vimos}. Within a projected separation of $1$~Mpc and $|\Delta v| = 500$~{\kms} of the absorber ($z = 0.83718$), eight galaxies are identified. The galaxy information is tabulated in Table \ref{tab:Table 2a} and Table \ref{tab:Table 2b}, and the distribution of galaxies is shown in Figure \ref{fig:galaxydist}. Of these, the 3 galaxies from the MUSE survey (G1, G2, and G3) are within $270$~kpc of the absorber, with the rest from the VVDS and Magellan surveys being further away. The MUSE and $HST$/ACS broad-band images of $1^{\prime}~\times~1^{\prime}$ encompassing these three galaxies is shown in Figure \ref{fig:galaxyimage}, and the MUSE spectra of the galaxies in Figure \ref{fig:galaxyspectra}. Over a wider projected separation of $5$~Mpc and $|\Delta v| \leq 1000$~{\kms}, 17 additional galaxies are identified. The VVDS survey has a magnitude limit of $i \lesssim 23$ which corresponds to $L \gtrsim 0.2L^*$ estimated using the Schechter luminosity function parameters given in \citet{2005ApJ...631..126D} at $z \sim 0.8$. To estimate the random probability of finding this many galaxies, we sampled 100 random regions within the VVDS survey area of 512 sq.arcmin, with a similar search window around $z = 0.837$. We found a less than $24$\% probability of finding more than 10 galaxies, and a less than $6$\% probability of finding more than 15. Thus, the line of sight is consistent with probing a galaxy over-density region at the absorber redshift, perhaps the intragroup medium in a galaxy group environment.

\renewcommand{\thefigure}{7}
\begin{figure*}
   \includegraphics[scale=0.54]{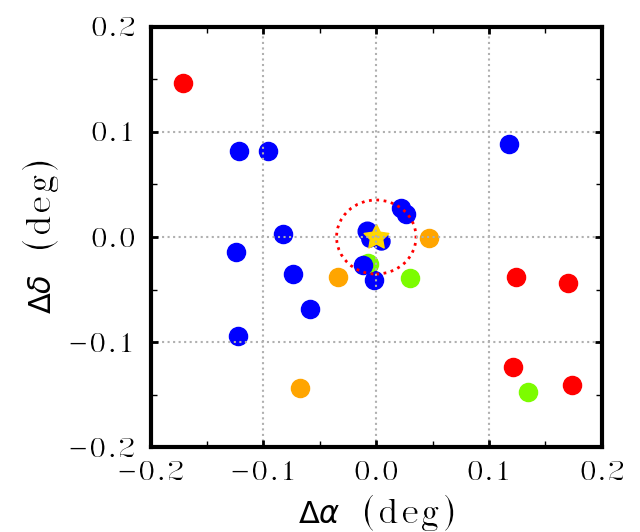}
   \includegraphics[scale=0.54]{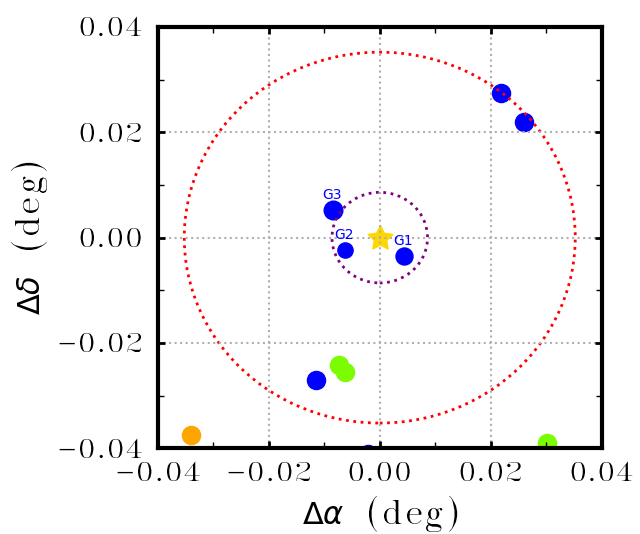}
   \caption{The \textit{Left} panel shows the large-scale distribution of galaxies within $|\Delta v| = 1000$~{\kms} and $5$~Mpc of projected separation from the absorber and the \textit{Right} panel is a zoom-in to the inner $|\Delta v| = 250$~{\kms} and $1$~Mpc. The quasar line of sight is indicated by the \textit{star} symbol at the center, and the galaxies are marked by the filled circles. The color coding corresponds to velocity offsets between the systemic redshifts of the galaxies and the absorber redshift of $z = 0.83718$. The \textit{blue, green, orange} and \textit{red} points are galaxies within $250$~{\kms}, between $250 - 500$~{\kms}, $500 - 750$~{\kms} and $750 - 1000$~{\kms} respectively. The two \textit{dashed} circles correspond to projected separations from the line-of-sight of $250$~kpc and $1$~Mpc.} 
    \label{fig:galaxydist}
\end{figure*}

\renewcommand{\thefigure}{8a}
\begin{figure*}
   \includegraphics[scale=0.55]{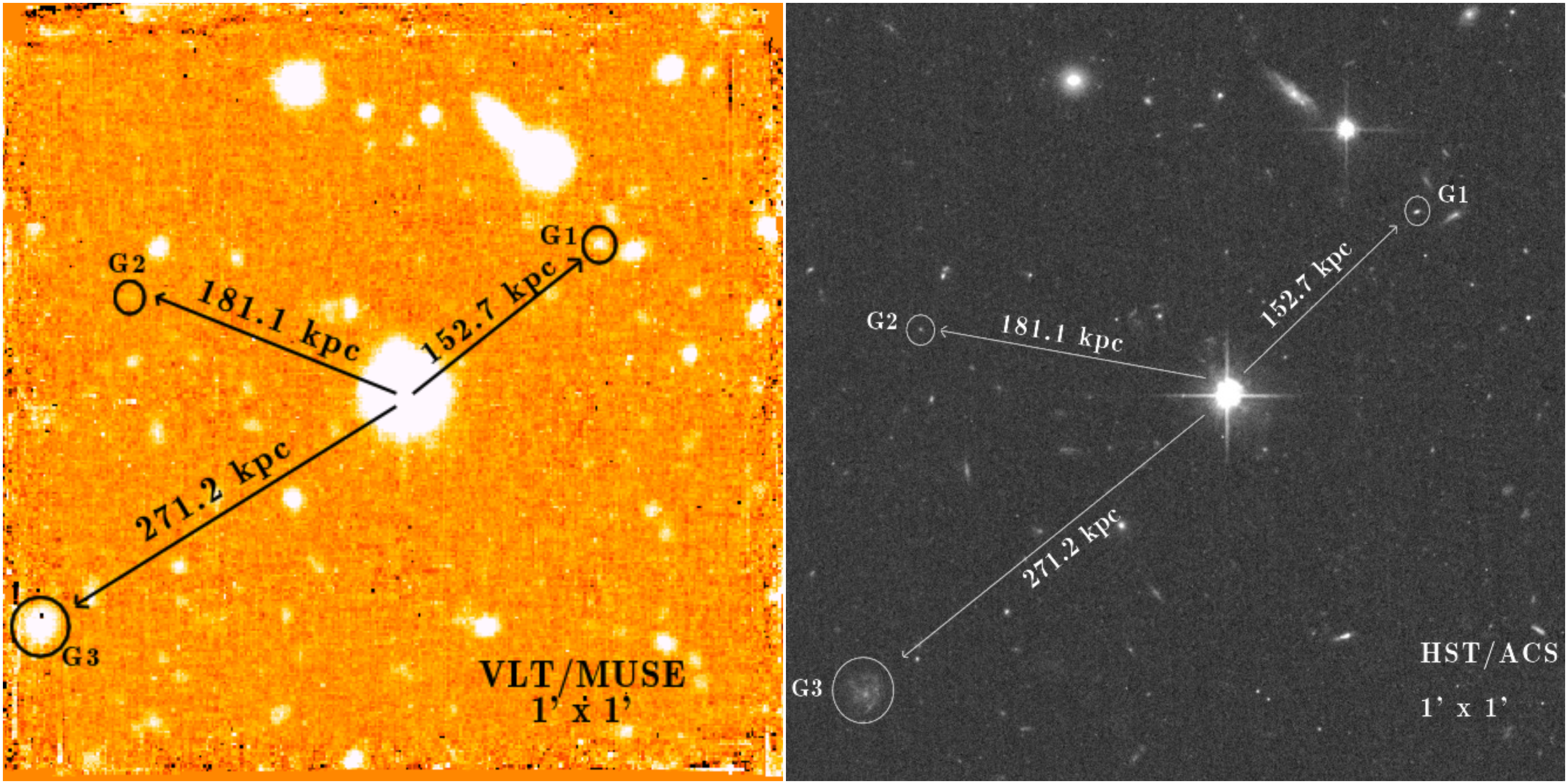}
   \caption{The VLT/MUSE (left panel) and HST/ACS (right panel) images of the $1^{\prime}~\times~1^{\prime}$ field centered around the quasar HE~$1003+0149$. The MUSE image was generated by taking the median of all the wavelengths from the MUSEQuBES data. The archival ACS image (program ID: 14269) was obtained using the broad-band filter F184W with central wavelength $8045$~{\AA} . The three galaxies closest to the absorber in projected separation are labeled. The galaxies are all within $125$~{\kms} of the absorber redshift, and possess stellar masses in the range of $M_* \approx 10^{9}, 10^{8.2}, 10^{9.8}$~M$_\odot$, and virial radii of $R_\mathrm{{vir}} \approx 85 - 115$~kpc, at normalized impact parameters of $\rho/R_\mathrm{{vir}} = 1.8, 3.0, 2.3$ for galaxies G1, G2, and G3 respectively. The detailed properties of the galaxies are given in Table \ref{tab:Table 2a}.} 
    \label{fig:galaxyimage}
\end{figure*}

\renewcommand{\thefigure}{8b}
\begin{figure*}
   \centering
   \includegraphics[scale=0.55]{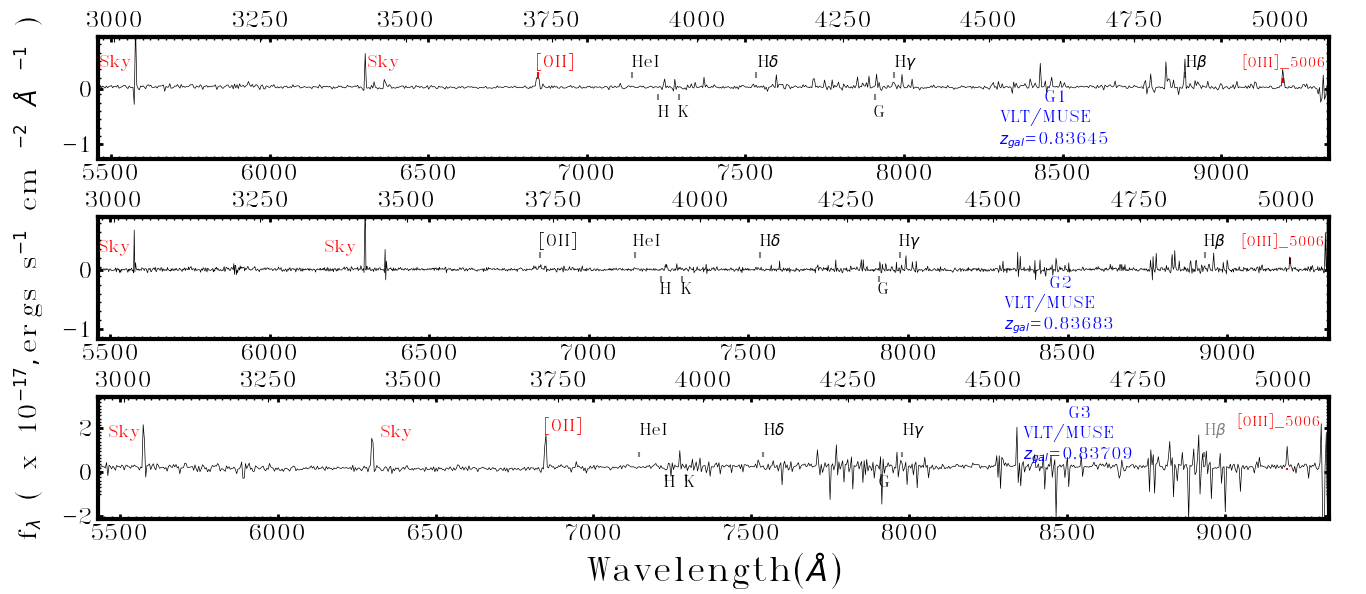}
   \caption{The $VLT$/MUSE spectra of galaxies G1, G2, and G3 shown in Figure \ref{fig:galaxyimage}, with the expected locations of prominent nebular emission lines labeled.
   The redshifts of the galaxies, and their SFRs were estimated using the [\OII]~$3727.092$~{\AA} and [\OIII]~$5008.240$~{\AA} emission lines. The former line is a feeble detection in G2. The horizontal axis at the bottom is observed wavelength and the top axis represents wavelength in the rest-frame of the galaxy.}
    \label{fig:galaxyspectra}
\end{figure*}

The galaxies G1, G2, and G3, which are closest to the absorber in projected separation (see Figure \ref{fig:galaxyimage}) , are faint dwarf galaxies with stellar masses in the range of $M_* = 10^{8.15} - 10^{9.80}$~M$_\odot$, and star-formation-rates of $0.1 - 1$~M$_\odot$~yr$^{-1}$. Numerous absorber-galaxy surveys have correlated high {\HI} column density with the CGM of galaxies  \citep{wakker2009relationship,prochaska2017cos}.The normalized projected separations of $\rho/R_{vir} \approx 1.8,~3.0,~2.3$ for G1, G2, and G3 galaxies indicate that the absorber is tracing gas outside of their virial radii. Here $R_{\mathrm{vir}}$ refers to $R_\mathrm{200}$, the radius where the mean enclosed density is $200$ times the critical density for that redshift. 

By analyzing QSO sight-lines targeted at foreground star-forming dwarf galaxies, \citet{2017wprb.confE..17J} found dwarf galaxy halos having a very small covering fraction for metal absorption compared to more massive galaxies, especially for impact parameters of $d/R_\mathrm{{vir}} > 1$. However, their sample was exclusively of field dwarf galaxies. Environmental effects such as tidal interactions, ram pressure stripping and mergers between galaxies, which are common in galaxy over-density regions, can significantly extend the cross-section of gas around galaxies \citep{domainko2006enrichment,2007PhDT........37H,2013MNRAS.431L...1S}, although within the virial halo, group galaxies tend to show a lower incidence of metal absorption compared to their field counterparts \citep{burchett2016deep,pointon2017impact}. Filamentary columns of gas, and spatially extended emission from galaxies, likely associated with stripped gas, have been detected in galaxy group environments \citep{Johnson_2018,Chen_2020}. The gas within galaxy groups may also tend to be strongly multiphase due to the frequent injection of hot gas from AGNs and star-forming galaxies raising the internal energy of the intragroup plasma.  From these it follows that the absorber complex has a higher likelihood of tracing the multiphase intragroup medium than gas bound to the halo of one of the nearest galaxies. 

Though the average metallicity of the absorber as obtained from the ionization models is low ($\sim 1/10$-th solar) and comparable to those measured for intra-group gas from X-ray emission \citep{helsdon2000intragroup}, the sub-solar [C/O] abundance for components 1 and 2 hints at chemical enrichment history by core collapse supernovae. The [C/$\alpha$] $< 0$ is a trend that is seen in a number of pLLS and LLS systems \citep{2016ApJ...833..283L}, as well as in lower column density gas \citep{Aguirre2008}. The abundance ratio serves as an aid to understand the chemical enrichment history of the gas. The dominant source of oxygen in the ISM is core-collapse supernovae involving $\geq 8$~M$_{\odot}$ stars \citep{iwamoto1999nucleosynthesis,nomoto2006nucleosynthesis}. On the other hand, carbon's pathway into the ISM is primarily through mass loss during the AGB phase of intermediate mass stars ($\approx 1 - 8$~M$_{\odot}$). The different time scales of these processes imply that the [C/O] ratio would tend towards solar only at late stages of ISM enrichment following star formation.

This leads to four possibilities. The first is where the absorber is tracing metals released into the intra-group medium from one or more of the nearby galaxies in early outflows triggered by correlated supernova events. Assimilation of these metals by the metal poor intragroup gas would result in the line of sight integrated metallicities coming out as low, despite the relative elemental abundances suggesting a Type II SNe enrichment pattern. Indeed, hydrodynamic simulations find Type II supernova driven outflows as necessary for reproducing not just the intragroup metallicities \citep{2016MNRAS.456.4266L}, but also the [O/Fe] abundance in them \citep{2008MNRAS.391..110D}. 

An alternative is for the gas to have been displaced from the nearby dwarf galaxies through tidal forces or hydrodynamic interactions between galaxies and the warm-hot ($T \gtrsim 10^5$~K) intra-group medium, if the line of sight is tracing a loose group.  The mass-metallicity relation for dwarfs implies baseline metallicities of [O/H] $\approx -0.9$ for galaxies with stellar masses around $M* \sim 10^8 - 10^9$~M$_\odot$ \citep{2017A&A...606A.115H}. Though this relationship has considerable scatter ($\approx \pm~0.5$~dex) in both stellar mass and ISM metallicity, the average metal abundance in the ISM being significantly sub-solar even when there is a star-formation history implies that dwarf galaxies retain less metals \citep{kirby2011metals,emerick2018metal}. The line of sight could be tracing such inherently low-metallicity ISM gas stripped through gravitational tidal forces between member galaxies or between a galaxy and the warm intragroup gas. \citet{2020MNRAS.497..498C} in their recent CUBS survey of LLS at $z < 1$ have hypothesized such an origin from dwarf galaxies for one of the absorbers in their sample. The absorber is near to a pair of galaxies of $M* \sim 10^8$~M$_\odot$, and $1/10$-th solar metallicity, but much closer in impact parameter $\rho < 40$~kpc than the dwarf galaxies of our sample. 

The third possibility is for the absorber to be tracing metal-poor gas streaming into the galaxy group environment from the cosmic web of inter-galactic filaments \citep[e.g.,][]{2016MNRAS.461..412Z}. Such high N(\HI) gas flows are seen in cosmological simulations \citep[e.g.,][]{2015MNRAS.453.4051E,2016MNRAS.461..412Z} and through observations of {\Lya} clouds in galaxy over-density spaces \citep{2012ApJ...754...84Y,muzahid2017}. Such gas defined as cold streams is yet to be shock heated to the virial temperatures of the dark haloes of galaxy groups/clusters. \citet{muzahid2017} detected {\HI} absorbers with $\log [N(\HI)/\cmsq] \approx 16.5 - 18.6$ in the outskirts [impact parameter, $\rho_{\mathrm{cl}} \approx (1.6 - 4.7)r_{500}$] of three galaxy clusters at $z \approx 0.46$. The temperatures inferred from the {\HI} and metal lines in those instances were consistent with photoionized gas. Based on ionization models, \citet{2019MNRAS.488.5327P} inferred that these absorbers were probing a cooler phase of intracluster gas different from the hot X-ray emitting plasma that dominates cluster cores with a plausible origin similar to the scenarios already discussed. In the case of the absorber clouds featured here, such an origin would also require a mixing of metals with the IGM from prior supernova enriched outflows.  

Finally, we cannot exclude the possibility of the absorber being directly associated with a very faint dwarf galaxy, much closer in impact parameter and below the detection threshold of the MUSE data. Using cosmological simulations, \citet{rahmati2014predictions} have shown that a great number of LLS could be linked to the virial haloes ($\rho \lesssim R_{200}$) of such faint galaxies with low stellar and gas masses ($M_* < 10^8$~M$_\odot$, $M_{\HI} < 10^8$~M$_\odot$), and little star-formation activity (SFR $\lesssim 10^{-2}$~M$_\odot$~yr$^{-1}$). The higher number densities of such low-mass, low-luminosity dwarfs in galaxy groups and clusters increases the odds of this scenario. 

\section{Conclusions}

We have undertaken a study of the galaxy environments and the properties of metal lines associated with three intervening {\HI} absorbers at closely separated redshifts of $z = 0.83697, 0.83745,$ and $0.83943$ corresponding to $|\Delta v| \approx 400$~{\kms} in line-of-sight velocity separation. The absorption line measurements for the three clouds, which include all the consecutive ionization stages from {\OI} to {\OV}, are summarized in Table~\ref{table1}. The key results are:

\begin{enumerate}

    \item The quasar field covered shows eight galaxies within a projected separation of  $1$~Mpc and $|\Delta v| = 500$~{\kms} of the absorber ($z = 0.83718$). Three among these are detected by MUSE to be within 270 kpc of the absorber, and are dwarf systems with $M_* \approx 10^8 - 10^9$~M$_\odot$, and SFR $\approx 0.5 - 1$~M$_\odot$~yr$^{-1}$. The normalized impact parameters of $\rho/R_{\mathrm{vir}} \approx 1.8, 3.0, 2.3$ for these three closest galaxies indicate the absorber complex probes gas beyond their individual virial halos. The VVDS and Magellan galaxy redshift surveys show 16 additional galaxies over a wider projected separation of $5$~Mpc and $|\Delta v| = 1000$~{\kms}.  The line-of-sight appears to be tracing the multiphase medium within a group of galaxies at the redshift of the absorber. We summarize the galaxy information in tabulated form in Tables~\ref{tab:Table 2a} and \ref{tab:Table 2b}.  

    \item  Photoionization modeling of the absorber complex reveals two phases, a low-ionization phase and a high-ionization phase. The {\OV} is found to trace higher-ionization gas compared to {\OII} and {\OIII}. For the low phase, the gas densities, $n_{\H}$, range between 10$^{-2.8}$--10$^{-1.9}$ cm$^{-3}$, the temperatures are in the range 10$^{4.3}$-10$^{4.7}$ K, and the metallicities range between $\sim$ 1/100th and 1/10th solar. For the high phase, densities are lower, but the properties are not as robustly constrained. The ionization modeling results are summarized in Table~\ref{tab:Table 3}. The marginalized posterior distributions for the metallicity ($\log (Z/Z_{\odot})$) and ionization parameter ($\log U$) for the low phase and high phase in components 1, 2, and 3 are presented in Figures~\ref{fig:corner1}, \ref{fig:corner2}, and \ref{fig:corner3}. The models are also used to predict the expected {\OVI} absorption associated with this absorber complex, for which there is no spectral coverage in the current data. The presence of two phases as revealed by component-by-component, multiphase Bayesian ionization modeling shows the absorption from kinematically coincident components to be from gas of different properties. 
    
    \item An inferred [C/O] $= -0.25^{+0.21}_{-0.69}$, and $-0.24^{+0.21}_{-0.82}$ in the components corresponding to the pLLS suggests a chemical enrichment history dominated by core collapse supernovae. This result is consistent with the trend seen in a number of pLLS and LLS systems~\citep{2016ApJ...833..283L}. In section~\ref{sec:discussion}, we discuss possible scenarios that can explain a sub-solar abundance in intragroup gas. 
    
\end{enumerate}

\section*{Data Availability}

The QSO spectroscopic data underlying this article can be accessed from the HST Spectroscopic Legacy archive (https://
archive.stsci.edu/hst/spectral\_legacy/) which is available in the public domain. The VVDS and VLT/MUSE galaxy data are available in VIMOS VLT Deep Survey (VVDS) database through http://cesam.lam.fr/vvds/, and the ESO Science Archive Facility (http://archive.eso.org/cms.html) respectively. The Magellan galaxy redshift survey data were provided by Dr.Hsiao-Wen Chen and Dr. John Mulchaey by permission. Data will be shared on request to the corresponding author with their permission.

\section*{Acknowledgements}

We thank Dr.Hsiao-Wen Chen and Dr. John Mulchaey for providing access to Magellan galaxy redshift data. We acknowledge the work of people involved in the design, construction and deployment of the COS on-board the Hubble Space Telescope and extend thanks to all those who had carried out data acquisition for the sight-line mentioned in this paper. We also thank the anonymous referee for carefully reviewing the manuscript and providing valuable comments. This research has made use of the HSLA database, developed and maintained at STScI, Baltimore, USA, and data from the VIMOS VLT Deep Survey, obtained from the VVDS database operated by Cesam, Laboratoire d'Astrophysique de Marseille, France. AN acknowledges support for this work from SERB through grant number EMR/2017/002531 from the Department of Science \& Technology, Government of India. SM thanks the Alexander von Humboldt Stifftung (Germany). This work is partly funded by Vici grant 639.043.409 from the Dutch Research Council (NWO).

\newcolumntype{Y}{>{\centering\arraybackslash}X}
\newpage
\begin{table}\centering
\setlength{\tabcolsep}{1.5pt}
\renewcommand{\arraystretch}{1.5}
\renewcommand\thetable{1}
\caption{Rest-frame equivalent widths, column densities and Doppler $b$-parameters for the three clouds that form the absorption complex at $z = 0.83718$ estimated using the AOD method and through Voigt profile fits. The {\MgII} and {\FeII} lines are covered by UVES, and the rest of the lines by COS. The equivalent width and integrated apparent column densities for Clouds 1 and 2 are combined. The last column corresponds to the velocity range of integration in the case of the AOD method and centroid of the component for each VP fit.} 
\label{table1}

\begin{tabularx}{\linewidth}{@{}p{4em}YYYY@{}}

\toprule
\multicolumn{1}{l}{Line} & \multicolumn{1}{r}{$W_r~(m\angstrom$)} &  \multicolumn{1}{c}{$\log [N/{\cmsq}]$} & \multicolumn{1}{c}{b (\kms)} & \multicolumn{1}{r}{v (\kms)} \\ 
\midrule

${\HI}~972$ & $547~\pm~24$  & $> 15.6$   & ... & $[-150,150]$ \\ 
            & $417~\pm~31$  & $> 15.5$   & ... & $[200, 500]$ \\ 
${\HI}~949$ & $512~\pm~12$  & $> 16.1$   & ... & $[-150,150]$ \\ 
            &  $325~\pm~16$  & $> 15.8$   & ... & $[200, 500]$  \\ 
${\HI}~937$ & $635~\pm~10$  & $> 16.4$   & ... & $[-150,150]$  \\ 
            & $270~\pm~11$  & $> 16.0$   & ... & $[200, 500]$ \\ 
${\HI}~930$ & $434~\pm~11$  & $> 16.4$   & ... & $[-150,150]$  \\
            & $209~\pm~9$   & $> 16.1$   & ... & $[200, 500]$ \\ 
${\HI}~926$ & $405~\pm~11$  & $> 16.6$   & ... & $[-150,150]$ \\
            & $191~\pm~10$  & $16.12~{\pm}~0.07$   & ... & $[200, 500]$ \\ 
${\HI}~923$ & $354~\pm~7$  & $> 16.6$   & ... & $[150, 450]$ \\
            & $137~\pm~9$   & $16.08~{\pm}~0.08$   & ... & $[-100, +100]$ \\ 
${\HI}~920$ & $323~\pm~11$  & $> 16.7$   & ... & $[150, 450]$  \\
            & $121~\pm~9$  & $16.13~{\pm}~0.06$   & ... & $[-100, +100]$ \\ 
${\HI}~919$ & $314~\pm~12$  & $> 16.8$   & ... & $[150, 450]$  \\
${\HI}~918$ & $269~\pm~12$  & $16.77~\pm~0.07$ & ... & $[-150,150]$  \\
${\HI}~917$ & $242~\pm~12$  & $16.79~\pm~0.06$ & ... & $[-100,100]$ \\
${\CII}~903.9$  & $< 54$  & $< 13.4$  & ... & $[-100,100]$  \\
                & $< 39$  & $< 13.2$  & ... & $[350,450]$  \\
${\CII}~903.6$  & $< 57$  & $< 13.7$  & ... & $[-100,100]$  \\ 
                & $< 39$  & $< 13.5$  & ... & $[350,450]$  \\
${\CIII}~977$   & $293~\pm~34$ & $13.96~\pm~0.37$ & ... & $[-200,100]$ \\
                & $179~\pm~32$ & $13.72~\pm~0.55$ & ... & $[250,450]$ \\
${\OI}~971$		& $< 99$ & $< 14.9$		 & ...		   & $[-100,100]$ \\
${\OI}~971$		& conta.	 & conta.		 & ...		   & $[250, 450]$ \\			
${\OII}~834$    & $52~\pm~8$ & $13.86~\pm~0.08$ & ... & $[-100,100]$ \\ 
                & $< 14$ & $< 13.2$ & ... & $[250, 450]$ \\ 

${\OIII}~702$   & $152~\pm~12$ & $14.59~\pm~0.10$ & ... & $[-150,150]$ \\ 
${\OIII}~832$   & $182~\pm~8$  & $14.60~\pm~0.13$ & ... & $[-150,150]$ \\
                & $117~\pm~11$ & $14.32~\pm~0.08$ & ... & $[300, 450]$ \\
${\OIV}~787$    & $135~\pm~9$  & $14.44~\pm~0.04$ & ... & $[-150,150]$ \\
${\OV}~629$     & $58~\pm~18$ &  $13.59~\pm~0.22$ & ... & $[-150,150]$ \\ 
                & $170~\pm~19$ &  $14.19~\pm~0.19$ & ... & $[250,500]$ \\ 
${\NeVIII}~770$ & $< 13$ & $< 13.4$ & ...  & $[-100,100]$ \\
                & $< 55$ & $< 14.1$ & ...  & $[250,450]$ \\
${\NeVIII}~780$ & $< 15$ & $< 13.8$ & ...  & $[-100,100]$ \\
${\MgX}~624$ & $ < 28$ & $< 14.3$  & ...  & $[-100,100]$ \\
             & $ < 35$ & $< 14.4$  & ...  & $[250,450]$ \\
\bottomrule
\end{tabularx}
\end{table}

\newcolumntype{Y}{>{\centering\arraybackslash}X}

\begin{table}\centering
\renewcommand\thetable{1}
\setlength{\tabcolsep}{1.5pt}
\renewcommand{\arraystretch}{1.5}
\caption{Continuation...}

\begin{tabularx}{\linewidth}{@{}p{4em}YYYY@{}}

\toprule
\multicolumn{1}{c}{Line} & \multicolumn{1}{c}{$\mathit{W_{r}}(m\AA)$} &  \multicolumn{1}{c}{$\log [N/{\cmsq}]$} & \multicolumn{1}{c}{b (\kms)} & \multicolumn{1}{r}{v (\kms)}  \\ 
\midrule
${\SIV}~748$ & $< 11$  & $< 12.7$  & ...  & $[-100,100]$ \\
             & $< 14$ & $< 12.8$  & ...  & $[250,450]$ \\
${\SV}~786$ & $< 10$   & $< 12.1$  & ...  & $[-100,100]$ \\
            & $< 11$ & $< 12.1$  & ...  & $[250,450]$ \\
${\SVI}~944$ & $< 18$ & $< 13.1$ & ...  & $[-100,100]$ \\
             & $ < 21$ & $< 13.1$  & ...  & $[250,450]$ \\
${\NIV}~765$    &  $< 33$  & $< 13.0$ & ... & $[-100,100]$ \\
${\MgII}~2796$  & $< 180$ & $< 12.6$ & ... & $[-100, 100]$ \\
${\MgII}~2796,2803$ &   ... &   $11.82~\pm~0.18$    & ... & $-34$ (from CCC-II) \\
${\FeII}~2600$ & $207$ & $< 13.2$ & ...  &  $[-100,100]$ \\ 
\hline
${\HI}~\small{972-916}$ & ... & $16.56~\pm~0.08$ & $26~\pm~4$ & $-37~\pm~6$ \\ 
                & ... & $16.42~\pm~0.10$ & $21~\pm~3$ & $31~\pm~4$ \\
                & ... & $16.14~\pm~0.07$ & $33~\pm~3$ & $369~\pm~7$ \\             
${\OII}~834$    & ... & $13.61~{\pm}~0.22$ & $21 \pm 5$ & $-34~\pm~2$\\ 
			    & ... & $13.72~\pm~0.19$ & $10~\pm~2$ & $32~\pm~2$ \\ 
${\CIII}~977$   & ... & $13.79~\pm~0.37$ & $23$ & $-46~\pm~4$ \\
			    & ... & $14.57~\pm~0.54$ & $10$   & $30~\pm~5$ \\
			    & ... & $13.64~\pm~0.20$ & $26~\pm~5$   & $357~\pm~5$ \\
${\OIII}~702, 832$ & ... & $14.43~\pm~0.07$ & $23~\pm~5$ & $-35~{\pm}~5$ \\ 
			       & ... & $14.45~\pm~0.18$ & $10~\pm~4$ & $31~{\pm}~3$ \\ 
			       & ... & $14.02~\pm~0.09$ & $26$ &    $372~{\pm}~3$ \\ 
${\OIV}~787$    & ...  & $14.17~\pm~0.11$ & $28~\pm~4$ & $-32~{\pm}~4$ \\ 
		        & ...  & $14.15~\pm~0.09$ & $15~\pm~2$ & $41~{\pm}~2$ \\
		        & ...  & $14.61~\pm~0.08$ & $30~\pm~3$ & $372^{\dagger}$ \\
${\OV}~629$     & ...  & $13.42~\pm~0.17$ & $23~\pm~5$ & $-39~{\pm}~9$ \\
                & ...  & $13.28~\pm~0.20$ & $12~\pm~4$ & $49~{\pm}~9$ \\ 
                & ...  & $14.26~\pm~0.13$ & $30~\pm~5$ & $402~{\pm}~9$ \\
\hline
\bottomrule
\end{tabularx}
  \begin{flushleft} $\dagger$ The {\OIV}~$787$ in component 3 is contaminated. The listed column density for this component is adopted as an upper limit.
	   \end{flushleft}
\end{table}

\begin{table*}
\renewcommand\thetable{2a}
\centering
	\caption{Galaxies within the $1^{\prime} \times 1^{\prime}$ FoV of MUSE and within $|\Delta v| = 250$~{\kms} from the absorber complex}.
	\begin{tabular}{cccccrrcr}
		\hline\hline
		Galaxy no. & RA (J2000)        &  Dec (J2000)      & $z_{\mathrm{gal}}$ & $\rho$~(kpc) & $|\Delta v|$~{\kms} & $R_\mathrm{{vir}}$~(kpc)    & SFR (M$_{\odot}$~yr$^{-1}$) & $M_*~(\mathrm{M_{\odot}})$ \\ \hline
		G1         & 151.392426 & 1.582795 & 0.83645  & 152.7  & -122.4 & $85.1$ & $0.45$    & $10^{9.0}$  \\
		           &  & & & & \\
		G2         & 151.403096 & 1.581628 & 0.83683  & 181.1  & -60.4  & $60.2$ & $0.11$    & $10^{8.2}$	 \\
		 &  & & & & \\
		G3         & 151.405204 & 1.574123 & 0.83709  & 271.2  & -17.9  &  $115.9$ & $1.07$    & $10^{9.8}$  \\
		 &  & & & & \\ \hline
        	\end{tabular}
	\label{tab:Table 2a}
	   \begin{flushleft} The $R_{\mathrm{vir}}$ estimated using the halo abundance matching relation of \citet{Moster13} can vary by a factor of $\approx 0.8$ due to the uncertainty in the best-fit values of the stellar mass to halo relation. Additionally, the stellar mass derived from SED fitting is typically uncertain by a factor of two.
	   \end{flushleft}
\end{table*}

\begin{table*}
\renewcommand\thetable{2b}
\begin{center}
	\caption{Galaxies over a wider projected separation from the absorber}
	\begin{tabular}{ccccrr}
		\hline\hline
		\multicolumn{6}{c}{Galaxies from the VVDS survey} \\
		\hline
    		RA (J2000)      &  Dec (J2000)      & $z_{\mathrm{gal}}$ & $z$-flag &    $\rho$~(kpc) & $|\Delta v|$~{\kms} \\ \hline
		151.40321       & 1.604889          & 0.83455   & 3 &   731.2   & -432.7    \\
		151.39899       & 1.620490          & 0.83690   & 3 &   1226.2  & -45.7     \\
        151.43079       & 1.616917          & 0.84143   & 3 &   1406.7  & 689.5     \\
        151.47066       & 1.614490          & 0.83720   & 3 &   2260.9  & 3.3       \\
        151.47998       & 1.576780          & 0.83690   & 2 &   2390.3  & -45.7     \\
        151.45576       & 1.647890          & 0.83660   & 3 &   2572.1  & -94.6     \\
        151.52089       & 1.593280          & 0.83610   & 4 &   3468.2  & -176.3    \\
        151.49249       & 1.497620          & 0.83850   & 3 &   3511.3  & 215.3     \\
        151.27310       & 1.617070          & 0.84200   & 4 &   3595.9  & 785.5     \\
        151.51836       & 1.497750          & 0.83660   & 3 &   4103.6  & -94.6     \\
        151.51927       & 1.673440          & 0.83590   & 3 &   4247.6  & -208.9    \\
        151.46417       & 1.723200          & 0.84030   & 1 &   4421.1  & 508.7     \\
        151.27534       & 1.702520          & 0.83130   & 2 &   4811.9  & -961.0    \\
        151.22653       & 1.623030          & 0.84220   & 2 &   4873.9  & 818.1     \\
        151.26263       & 1.726370          & 0.83960   & 3 &   5511.0  & 394.7     \\
        151.22346       & 1.720240          & 0.84300   & 3 &   6204.1  & 948.2     \\
        151.56778       & 1.432900          & 0.83250   & 1 &   6252.3  & -764.7    \\ \hline
        \hline
        \multicolumn{6}{c}{Galaxies from the Magellan survey} \\
        \hline
        RA (J2000)  &    Dec (J2000)      & $z_{\mathrm{gal}}$ & z-flag & $\rho$~($\mathrm{kpc}$)     & $|\Delta v|$~{\kms}      \\
        \hline
             151.40421 & 1.60364 & 0.8351 & - & 715.6 & -339.8 \\
             151.40832 & 1.60642 & 0.8370 &- & 831.1 & -29.4 \\
             151.37085 & 1.55757 & 0.8381 & -  & 963.2 & 150.2 \\
             151.37500 & 1.55198 & 0.8364 & -  & 993.9 & -127.4\\
             151.35056 & 1.58031 & 0.8406 & -  & 1327.0 & 557.9 \\
             151.36676 & 1.61830 & 0.8397 & -  & 1394.4 & 411.2\\
             151.27925 & 1.49063 & 0.8371 & -  & 4173.0  & -13.1\\
             \hline
    \end{tabular}
    	\label{tab:Table 2b}
    \end{center}
    \begin{flushleft} The galaxy information in this table is taken from the VVDS survey\footnote{http://cesam.lam.fr/vvds/index.php}. Column 4 is a spectroscopic reliability flag as mentioned in Le F\'evre et al. (2013) with the following probabilities: 4 represents $100$\% probability that the listed redshift is correct, 3 is $95-100$\%, 2 is $75-85$\%, and 1 is $50-75$\%. We note that the redshift errors for galaxies from the Magellan data are $\approx 60$~{\kms} (resolution of $R \approx 600 - 800$), whereas VVDS has $R \approx 200$ for VVDS, and higher redshift errors of $\approx 150 - 300$~{\kms}. We have quoted the redshifts to more number of digits for the VVDS data as exactly given in the VVDS catalogue.\end{flushleft}
\end{table*}

\begin{table*}
\renewcommand\thetable{3}
\begin{center}
\caption{\bf The gas phases in the $z = 0.83718$ absorber towards HE~$1003+0194$}
\begin{tabular}{ccrrrrrrr} \hline \hline
Phase & Optimizing Ion & v (\kms) & $\log (Z/Z_\odot)$ & $\log U$ & $\log [n(\H)/\cc]$& $\log (T/K) $ & $\log L$ (kpc) & $\log [N(\HI)/\cmsq]$\\ \hline 

\textcolor{blue}{Component 1 Low} & \textbf{\OII} & -34 & $-1.20^{+0.04}_{-0.04}$ & $-2.94^{+0.04}_{-0.04}$ & $-2.20^{+0.04}_{-0.04}$ & $4.29^{+0.01}_{-0.01}$ &  $-0.18^{+0.10}_{-0.13}$ & $16.51^{+0.03}_{-0.03}$\\
\textcolor{green}{Component 2 Low} & \textbf{\OII} & 32 & $-0.97^{+0.05}_{-0.05}$ & $-3.20^{+0.11}_{-0.13}$ & $-1.94^{+0.13}_{-0.11}$ & $4.21^{+0.03}_{-0.03}$ & $-1.02^{+0.30}_{-0.28}$ & $16.32^{+0.05}_{-0.05}$\\
\textcolor{red}{Component 3 Low} & \textbf{\OIII} & 372 & $-2.33^{+0.22}_{-0.21}$ & $-2.36^{+0.04}_{-0.02}$ & $-2.79^{+0.02}_{-0.04}$ & $4.71^{+0.05}_{-0.06}$ & $1.22^{+0.25}_{-0.19}$ & $16.06^{+0.03}_{-0.03}$\\
\textcolor{BurntOrange}{Component 1 High} & \textbf{\OV} & -39 & $>-2.45$ & $-1.74^{+0.68}_{-0.17}$ & $-3.40^{+0.17}_{-0.68}$ & $4.82^{+0.28}_{-0.72}$ & $-0.30^{+2.67}_{-1.57}$ & $<14.13$\\
\textcolor{BurntOrange}{Component 2 High} & \textbf{\OV} & 49 & $-2.03^{+0.50}_{-0.15}$ & $-2.11^{+0.10}_{-0.07}$ & $-3.03^{+0.07}_{-0.10}$ & $4.95^{+0.06}_{-0.08}$ & $1.44^{+1.44}_{-0.71}$ & $15.12^{+0.31}_{-0.57}$ \\
\textcolor{BurntOrange}{Component 3 High} & \textbf{\OV} & 402 &$>0.39$ & $-1.03^{+0.68}_{-0.41}$ & $-4.12^{+0.41}_{-0.68}$ & $4.65^{+0.87}_{-0.29}$ & $-0.24^{+2.98}_{-0.66}$ & $<13.17$\\

\hline 

\end{tabular} \\
\label{tab:Table 3}     
\end{center}
\begin{flushleft}
Properties of the different gas phases present in $z = 0.83718$ towards HE~$1003+0194$ traced by their respective optimizing ions. The marginalized posterior values with the median along with the upper and lower bounds associated with 95\% credible interval ($2\sigma$) are given. The component names are color coded to indicate the respective components in Figure~\ref{fig:systemplotmodel2}. The ions used for optimizing the models and their velocity centroids are listed in columns 2 and 3 respectively. The listed metallicities are based on fine-tuning the models such that their predicted column densities match the column density of the optimizing ion, alongside as many other ions as possible. Components 1, and 2 require [C/O] $\approx -0.24$~dex to match the observed {\CII} from the same phase as {\OII}.
\end{flushleft}
\end{table*}


\clearpage

\bibliographystyle{mnras}
\bibliography{main} 

\appendix
\onecolumn
\section{The Feature Corresponding to the {\SII}~$765$~{\AA} Line}
\label{Appendix A}

The absorption at $\lambda = 1406.6$~{\AA} with rest-frame equivalent width of $W_r = 42~\pm~11$~m{\AA} corresponds to the expected location of redshifted {\SII}~$765$ for components 1 and 2. There is no absorption seen at the expected location of the third component, down to a $3\sigma$ upper limit of $W_r < 33$~m{\AA}. We rule out the detection at $\lambda = 1406.6$~{\AA} as genuine {\SII}~$765$ for the following reasons. In Figure A1, we compare the apparent column density profiles of the presumed {\SII}~$765$ with the {\OII}~$834$, and {\OIII}~$832$ lines. The separation between the components in the supposed {\SII} line is $\Delta v \approx 45$~{\kms}, as opposed to $\approx 70$~{\kms} between the components seen for the other metal lines, and the higher order Lyman lines. The {\SIII}~$710, 681$, and {\SIV}~$657$ lines are also non-detections at the $3\sigma$ level. Furthermore, as shown in Figure A2, the two phase photoionization model that simultaneously explains the {\OII}, and {\OIII} ionization stages of oxygen, and the {\CIII}, contradicts the absorption presumed to be {\SII}~$765$, in spite of the creation and destruction energies of {\SII} being close to {\OII} (10.4 eV/23.3 eV vs. 13.6 eV/35.12 eV). For these reasons, we eliminate the absorption as unidentified contamination. A thorough search for all absorption lines along the sightline will be required to identify this contamination. 

\begin{figure*}
   \includegraphics[width=0.5\textwidth]{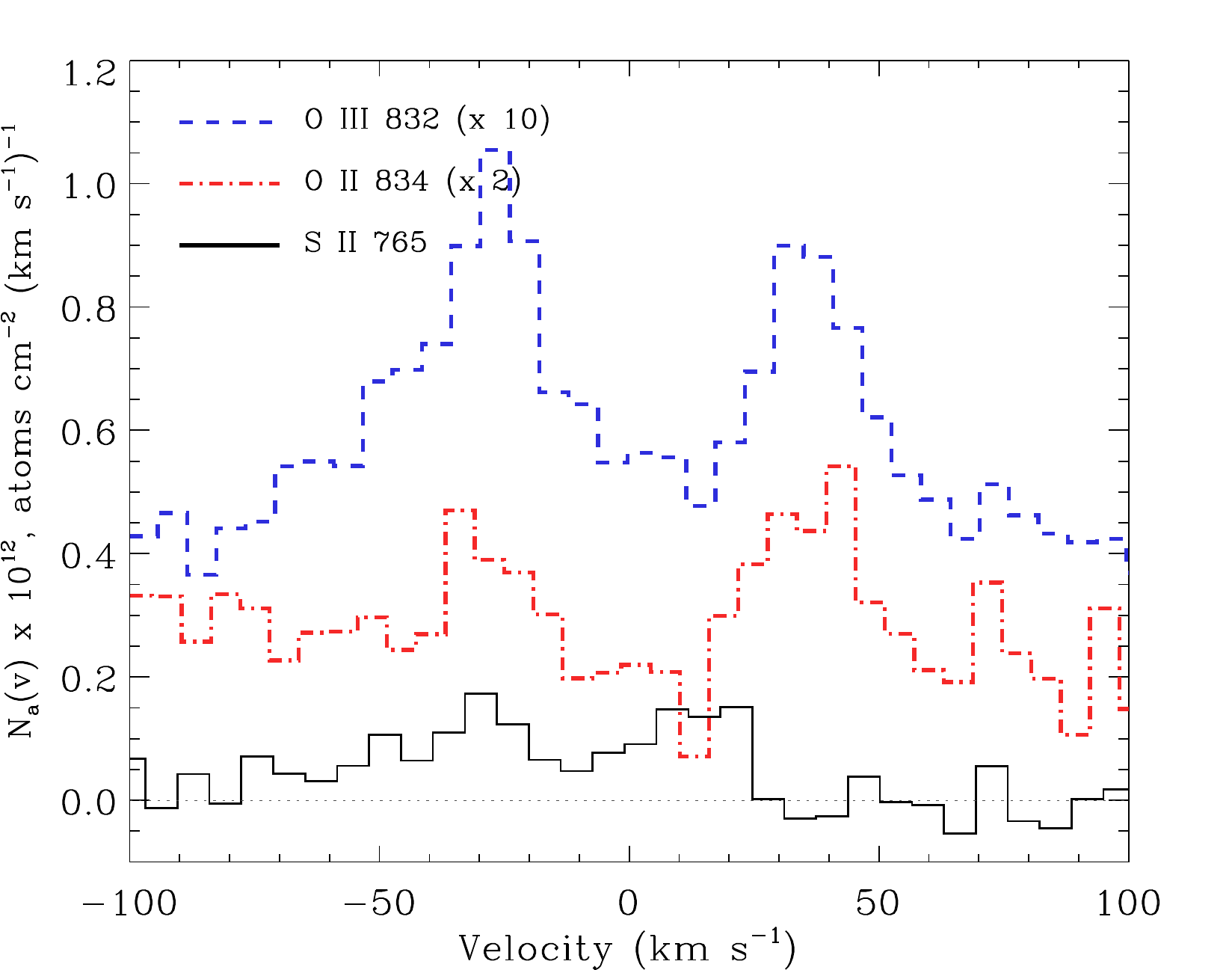}
   \caption{A comparison of the run of apparent column density with rest-frame velocity corresponding to $z = 0.83718$ for the putative {\SII}~$765$ feature, and the {\OII}~834 line. The ions have similar ionization energies and are therefore expected to trace gas of same phase. For comparison, the {\OIII} line is also included.  The profiles have been offset vertically for clarity. The components at the expected location of {\SII} have a different velocity separation compared to the other lines.} 
    \label{fig:appendixfig1}
\end{figure*}

\begin{figure*}
\begin{center}
	\includegraphics[width=0.6\textwidth]{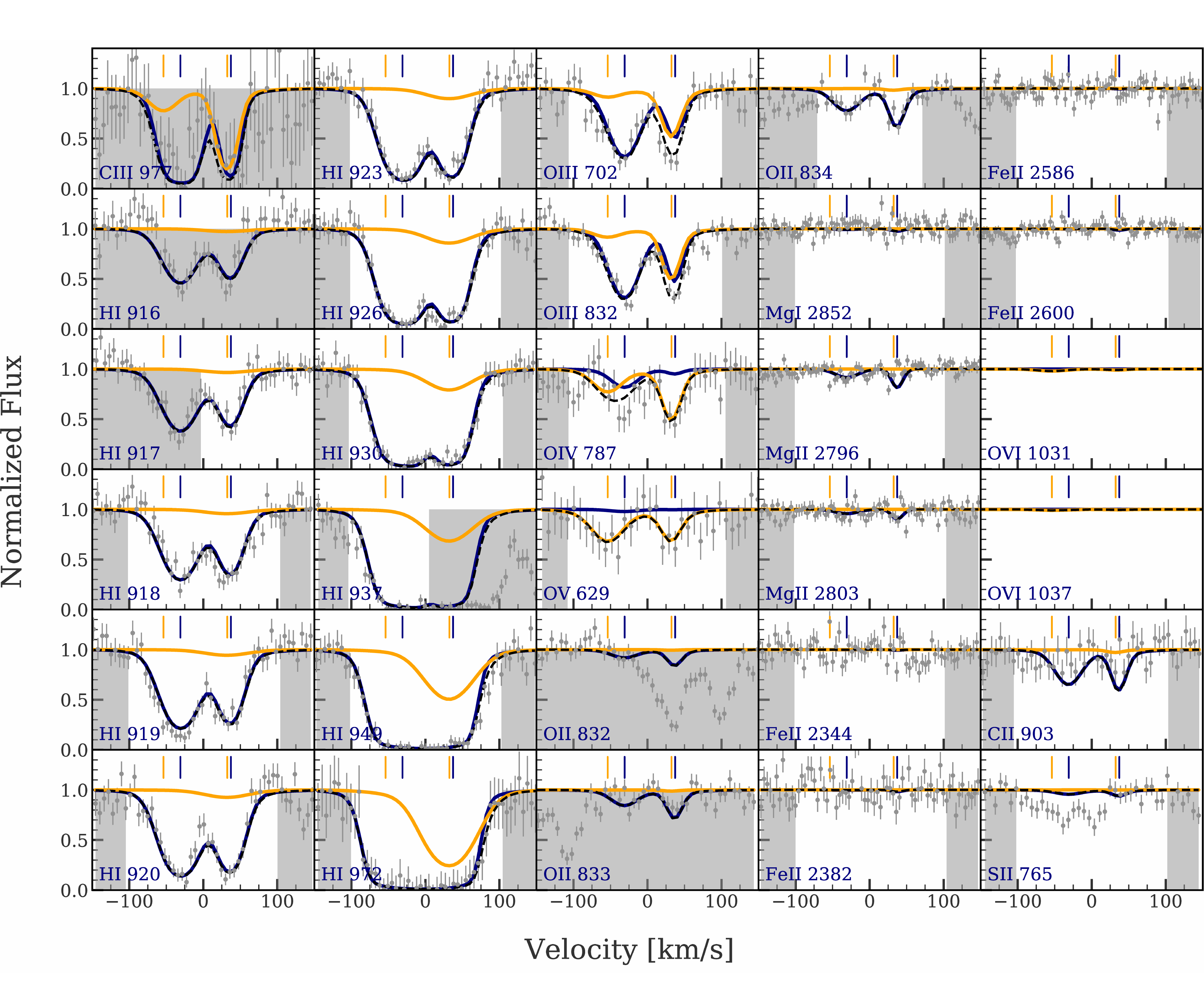}
    \caption{Photoionization models that simultaneously explain the two component absorption using a two phase model predict little {\SII}, which is inconsistent with the absorption seen at the expected location of {\SII}~$765$ (\textit{bottom right panel}), indicating that the feature is unrelated to the pLLS.}
    \label{fig:third_sightline}
\end{center}
\end{figure*}

\clearpage
\onecolumn
\section{Column Density Predictions from the Maximum Likelihood Estimation Model}
\label{Appendix B}

The Table \ref{tab:Table B1} shows the column density predictions for the {\HI} and the various metal ions shown in Figure~\ref{fig:systemplotmodel2} based on the maximum likelihood estimation model. For components 1 and 2, the ionization models were optimized on {\OII}. In component 3, {\OIII} was chosen as the optimizing ion, since {\OII} is a non-detection in this component. The table also lists the observed column densities or their upper limits (non-detections) for the three components in the last three columns.  

\begin{table}
\renewcommand\thetable{B1}
\begin{center}
\caption{Ionization model predictions for the ions shown in Figure~\ref{fig:systemplotmodel2}}
\begin{tabular}{lrrrrrrrrr}\hline \hline
Ion  & Comp1-Lo & Comp2-Lo & Comp3-Lo & Comp1-Hi & Comp2-Hi & Comp3-Hi  & Comp1-Obs & Comp2-Obs & Comp3-Obs \\ \hline 
{\HI}   & 16.51    & 16.29    & 16.06    & 13.78    & 15.21    & 12.41  & 16.56 & 16.42 & 16.14  \\
{\CII}  & 13.38    & 13.29    & 12.84    & 10.45    & 12.02    & 7.61   & < 13.4 & < 13.4 & < 13.2   \\
{\CIII} & 14.16    & 13.84    & 13.82    & 12.38    & 13.59    & 10.61  & 13.79 & 14.57 & 13.64  \\
{\OII}  & 13.61    & 13.72    & 12.53    & 11.36    & 12.26    & 8.04   & 13.61 & 13.72 & 13.79 \\
{\OIII} & 14.38    & 14.02    & 14.02    & 13.34    & 14.09    & 11.16  & 14.43 & 14.45 & 14.02  \\
{\OIV}  & 13.57    & 12.79    & 13.85    & 13.84    & 14.07    & 13.12  & 14.17 & 14.15 & 14.61  \\
{\OV}   & 12.06    & 11.08    & 12.89    & 13.42    & 13.28    & 14.26  & 13.42 & 13.28 & 14.26  \\
{\MgI}  & 10.28    & 10.52    & 6.08     & 3.46     & 4.56     & 3.16   & < 12.3 & < 12.3 & < 12.2  \\
{\MgII} & 11.86    & 11.92    & 8.73     & 7.03     & 7.83     & 4.81   & 11.82 & < 12.6 & < 12.6 \\ 
\hline 
\end{tabular}\\
\label{tab:Table B1}
\end{center}
\end{table}

\label{lastpage}

\end{document}